\documentclass[conference]{IEEEtran}
\IEEEoverridecommandlockouts
\usepackage{cite}
\usepackage{amsmath,amssymb,amsfonts}
\usepackage{algorithmic}
\usepackage{graphicx}
\usepackage{textcomp}
\usepackage{xcolor}
\usepackage{braket}
\usepackage{algorithm}

\usepackage{tikz}
\newcommand*\circled[1]{\tikz[baseline=(char.base)]{
            \node[shape=circle,draw,inner sep=0.3pt] (char) {#1};}}

\usepackage{wasysym}
\def\BibTeX{{\rm B\kern-.05em{\sc i\kern-.025em b}\kern-.08em
    T\kern-.1667em\lower.7ex\hbox{E}\kern-.125emX}}
\begin{document}

\title{O3LS: Optimizing Lattice Surgery via Automatic Layout Searching and Loose Scheduling
\thanks{$^*$Co-first authors}
\thanks{$^\dagger$Co-corresponding authors}
\thanks{This work has been partially supported by the National Key R\&D Program of China (Grant No.~2024YFB4504001), the National Natural Science Foundation of China (Grant Nos.~12447107, 62302395, and 62421002), the Fundamental and Interdisciplinary Disciplines Breakthrough Plan of the Ministry of Education of China (Grant No.~JYB2025XDXM202), the Aid Program for Science and Technology Innovative Research Teams in Higher Educational Institutions of Hunan Province, and the Guangdong Provincial Quantum Science Strategic Initiative (Grant Nos.~GDZX2403008 and GDZX2503001).}
}

\author{\IEEEauthorblockN{1\textsuperscript{st} Chenghong Zhu$^*$}
\IEEEauthorblockA{
\textit{The Hong Kong University of} \\ 
\textit{Science and Technology (Guangzhou)} \\
Guangzhou, China}
\and
\IEEEauthorblockN{2\textsuperscript{nd} Xian Wu$^*$}
\IEEEauthorblockA{
\textit{The Hong Kong University of Science} \\ 
\textit{ and Technology (Guangzhou)} \\
Guangzhou, China}
\and
\IEEEauthorblockN{3\textsuperscript{rd} Jiahan Chen}
\IEEEauthorblockA{
\textit{The Hong Kong University of} \\ 
\textit{Science and Technology (Guangzhou)} \\
Guangzhou, China}
\and
\IEEEauthorblockN{4\textsuperscript{th} Keming He}
\IEEEauthorblockA{
\textit{The Hong Kong University of} \\ 
\textit{Science and Technology (Guangzhou)} \\
Guangzhou, China}
\and
\IEEEauthorblockN{5\textsuperscript{th} Junjie Wu}
\IEEEauthorblockA{\textit{College of Computer Science and Technology} \\
\textit{National University of Defense Technology}\\
Changsha, China}
\and
\IEEEauthorblockN{6\textsuperscript{th} Xin Wang$^\dagger$}
\IEEEauthorblockA{
\textit{The Hong Kong University of} \\ 
\textit{Science and Technology (Guangzhou)} \\
Guangzhou, China \\ 
felixxinwang@hkust-gz.edu.cn}
\and
\IEEEauthorblockN{7\textsuperscript{th} Lingling Lao$^\dagger$}
\IEEEauthorblockA{\textit{College of Computer Science and Technology} \\
\textit{\centerline{National University of Defense Technology}}\\
Changsha, China \\
laolinglingrolls@gmail.com}
}

\maketitle

\begin{abstract}
Toward the large-scale, practical realization of quantum computing, quantum error correction is essential. Among various quantum error-correcting codes, the surface code stands out as a leading candidate, and lattice surgery based on surface codes has emerged as a promising technique for fault-tolerant quantum computation (FTQC). However, implementing quantum algorithms using lattice surgery introduces both resource and time overhead. Existing approaches typically focus on large layout designs, with compiler passes aimed primarily at optimizing time overhead. This often overlooks the trade-off between rotation bottlenecks and movement distance, which leads to inefficient resource utilization and prevents further reduction of the quantum computation failure rate.

To address these challenges, we introduce O3LS, a framework for optimizing lattice surgery through automatic layout search and loose scheduling. O3LS achieves an optimal balance by automatically generating squeezed data layouts to reduce space requirements and employing loose scheduling algorithms combined with circuit synthesis techniques to reduce time overhead, thereby effectively minimizing overall logical error rates.
Numerical results indicate that O3LS can reduce space overhead by 28.0\% over standard layouts and 46.7\% over sparse layouts without increasing the number of time steps, leading to suppression of logical error rates by up to 16\% relative to larger data layout designs. O3LS can also achieve time overhead reductions of 36.07\% and 24.76\% in compact and standard data layout designs,  respectively. It suppresses logical error rates by up to an order of magnitude compared to prior compilers that focus primarily on maximizing parallelism.
\end{abstract}

\begin{IEEEkeywords}
fault-tolerant quantum computation, surface code, lattice surgery, quantum compiler design
\end{IEEEkeywords}

\section{Introduction}

Quantum computers are expected to offer practical advantages over classical computers in solving certain classes of problems~\cite{shor1999polynomial, harrow2009quantum}. While noisy intermediate-scale quantum (NISQ) devices~\cite{preskill2018quantum} have enabled significant theoretical and experimental progress~\cite{arute2019quantum, cerezo2021variational, ebadi2022quantum, liu2025certified}, their performance remains limited by noise, hindering practical quantum advantage. This necessitates the use of quantum error correction (QEC)~\cite{shor1995scheme, gottesman1997stabilizer, fowler2012surface, bravyi2024high, yi2024complexity}, which is critical for enabling fault-tolerant quantum computing (FTQC). 

A series of experimental demonstrations have validated the feasibility of QEC~\cite{marques2022logical, zhao2022realization, krinner2022realizing, caune2024demonstrating, eickbusch2024demonstrating, wang2025demonstration}. Among the various QEC codes, the surface code has emerged as a leading candidate for realizing FTQC, owing to its relatively high error threshold, compatibility with nearest-neighbor qubit connectivity and various schemes for universal quantum computation. Notably, recent implementations on Google’s Willow processor have demonstrated exponential noise suppression using the surface code~\cite{acharya2024quantum, ai2024quantumwillow}, representing a significant milestone toward practical FTQC. 

A prominent technique for achieving universal quantum computation on 2D nearest-neighbor devices is lattice surgery, which implements logical multi-qubit gates by merging and splitting planar code patches~\cite{horsman2012surface,Litinski2019}. This approach enables the execution of logical operations within a 2D nearest-neighbor layout and introduces additional compilation challenges, particularly in ancilla routing and the design of data layouts, with the ultimate goal of minimizing the logical error rate (LER).

\begin{figure}[t]
    \centering
    \includegraphics[width=1\linewidth]{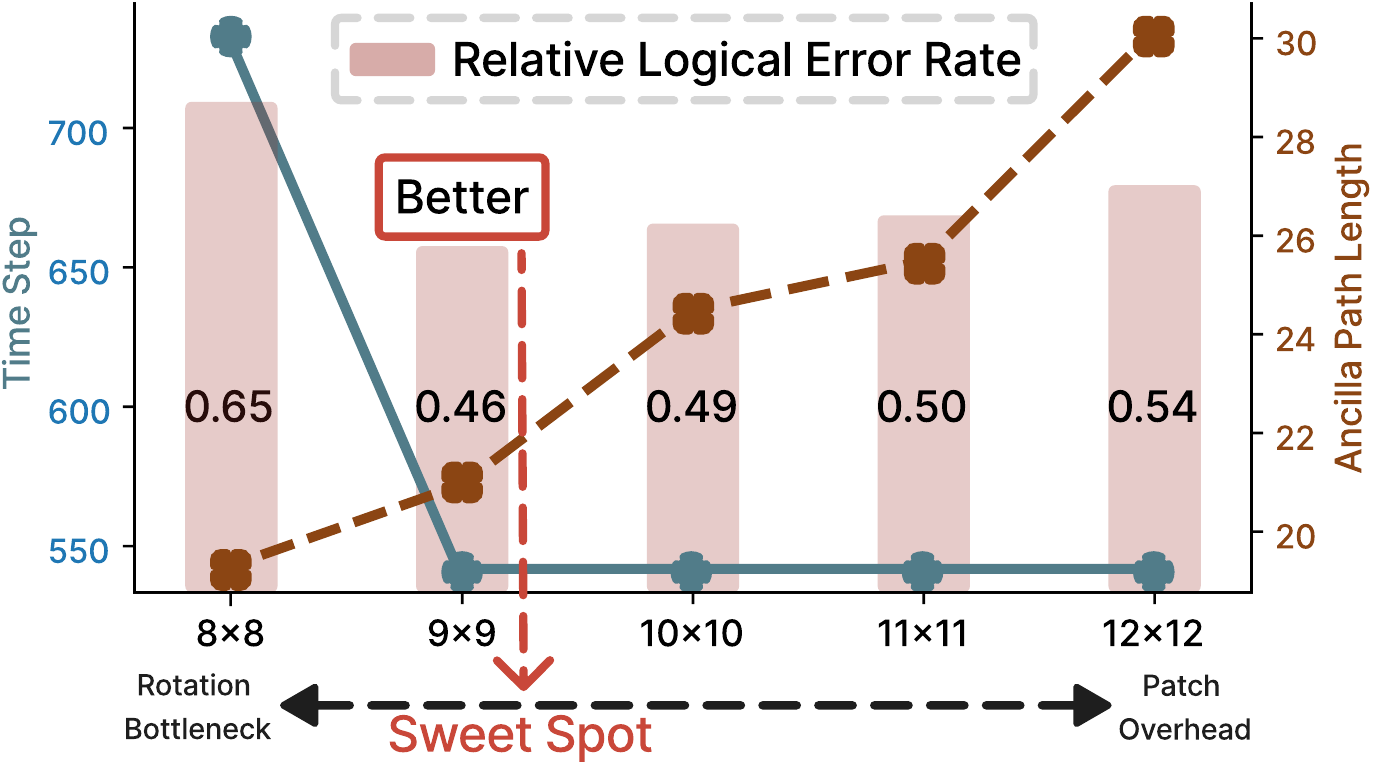}
    \caption{O3LS can achieve comparable time overheads by automatically designing squeezed data layouts, even compared to larger layouts (e.g. $10\times 10$ and $12\times 12$). Moreover, squeezed layouts can reduce ancilla patch distances, thereby leading to a lower logical error rate (sweet spot).}
    \label{fig:enter-label}
\end{figure}

Existing compilers~\cite{hirano2025localityaware, Tan2024LaS,ueno2024high} for lattice surgery compilation primarily focus on large or sparse data layouts to maximize parallelism. This strategy aims to reduce the total number of time steps required for executing quantum circuits, thereby minimizing LER. However, it overlooks a critical factor: \textbf{larger data layouts often result in increased movement distances}, which in turn elevate the likelihood of idle memory errors and ultimately degrade LER~\cite{kan2025sparo}. In contrast, compact data layouts can reduce the physical qubit footprint and limit movement overhead. Nevertheless, \textbf{if the layout architecture becomes too compact, it may significantly increase time costs} due to reduced parallelism. This reveals \textbf{a fundamental trade-off between the number of time steps and the movement distance} as illustrated in Fig.~\ref{fig:enter-label}, which must be carefully balanced to optimize overall LER.

Motivated by these insights, our work leverages data layout design (sweet spot in Fig.~\ref{fig:enter-label}) to reduce overall space overhead, while incorporating an optimized scheduling strategy to minimize time costs. Together, these approaches aim to achieve a more balanced and efficient reduction in the overall LER. In summary, our key contributions are:
\begin{itemize}
    \item We develop a lattice surgery compiler O3LS that integrates an automatic data layout design algorithm, an optimized synthesis algorithm, and a loose scheduling strategy to minimize both the space and time overhead.
    \item We propose an automatic logical qubit data layout design strategy aimed for squeezing data layouts, enabling more efficient use of limited qubit resources to support larger quantum applications.
    \item We design an optimized synthesis algorithm that improves Pauli operator cancellation, thereby reducing the time overhead of lattice surgery in more compact data layout designs and making it well-suited for integration with the data layout design algorithm.
    \item We also present a loose scheduling mechanism that dynamically reassigns patch functionalities. This approach reduces redundant patch movements and eliminates unnecessary operations.
\end{itemize}
The evaluation results demonstrate that O3LS simultaneously achieves time step reductions of 36.07\% and 24.76\%, and space overhead reductions of 28.0\% and 46.7\% on average, compared to previous compilers executed on fixed layouts. These improvements contribute to an effective suppression of overall logical error rates.

\section{Background and Motivation}

\subsection{Quantum Computing and Quantum Error Correction}

\textbf{Introduction to Quantum Computation.} 
In quantum computing, the basic unit of quantum information is qubit, which has two basis states, typically denoted as $\ket{0}$ and $\ket{1}$. A qubit can exist in a superposition expressed as $\ket{\psi} = \alpha\ket{0} + \beta\ket{1}$, where $\alpha,\beta\in \mathbb{C}$ and $|\alpha|^2 + |\beta|^2 = 1$. Quantum gates serve as fundamental operations for manipulating qubit states.

\textbf{Quantum Error Correction.} Quantum computers are vulnerable to noise that disturbs quantum states. Quantum error correction codes (QECCs) address this by encoding a logical qubit into multiple physical qubits.
Among these, the surface code~\cite{fowler2012surface} is particularly notable for its topological structure and compatibility with universal quantum operations. In Fig.~\ref{fig:surface_code_example}(a), we show an example of our FTQC scheme using a surface code tile. Each tile has four boundaries, which are either of X or Z type, indicated by white and red lines. The logical $Z$ operator is defined as the tensor product of physical $Z$ operators along a string of data qubits, as illustrated by the red dashed lines. The logical $X$ operator is defined similarly using physical $X$ operators.

\textbf{Common Gates and Their Decomposition.} The universal gate set Clifford+$T$ is well-suited for surface codes. It consists of the Hadamard gate $H$, the phase gate $S$, the $T$ gate requiring the consumption of magic state $\ket{0} + e^{i\pi/4}\ket{1}$, and the controlled-NOT gate ($\mathrm{CNOT}$). 
However, implementing $T$ gates requires the consumption of magic states, which must be generated through magic state distillation protocols~\cite{magicdistillation1, magicdistillation2, magicdistillation3, magicdistillation4} to enable universal quantum computation.

\subsection{Lattice Surgery in Surface Code}\label{lattice_surgery_preliminary}

\begin{figure}[h]
    \centering
    \includegraphics[width=1\linewidth]{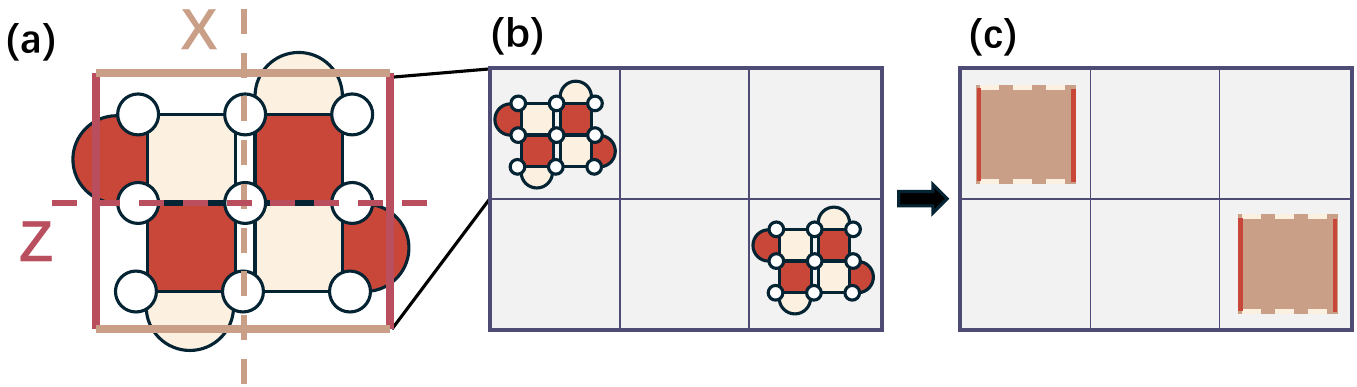}
    \caption{(a) Example of the distance-3 surface code. (b) Logical qubits are encoded by portions of a device’s physical lattice into patches. (c) Abstraction of logical qubits into patches, where the white dashed line represents the X operator and the red line represents the Z operator.}
    \label{fig:surface_code_example}
\end{figure}

\textbf{Abstraction of surface code to patches.} Each patch corresponds to a distance-$d$ surface code, which encodes a logical qubit using $d^2$ physical data qubits, as illustrated in Fig.\ref{fig:surface_code_example}(b) for the case of $d = 3$. These surface codes are then abstracted as patches placed on tiles, as shown in Fig.\ref{fig:surface_code_example}(c). In this abstraction, dashed boundaries represent $Z$ operators, while solid boundaries represent $X$ operators. 
These operators can be represented as edges on the patch and, for convenience, will be referred to as X- and Z-edges.
The key performance metric is the implementation of quantum algorithms using the minimum number of tiles and time steps—collectively referred to as the space-time volume. Here, the unit of time corresponds to round of code cycles, which we denote using symbol \clock.

\begin{figure}[t]
    \centering
    \includegraphics[width=0.9\linewidth]{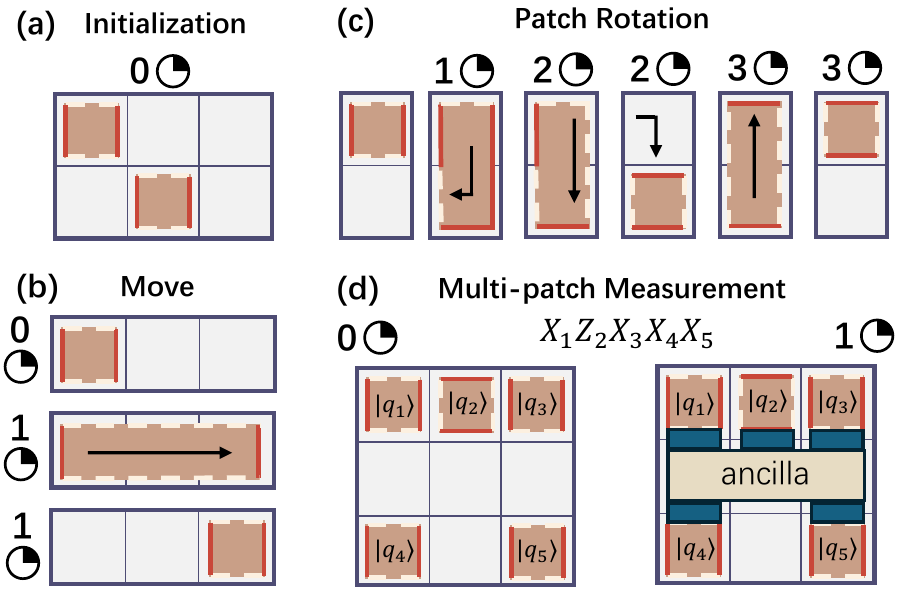}
    \caption{Patch operations and their related time costs. }
    \label{fig:rules_cost}
\end{figure}

\textbf{Patch Operations.} \textit{(a) Initialization.} A single-qubit patch can be initialized in the states $\ket{+}$ or $\ket{0}$ (Fig.~\ref{fig:rules_cost}a), and two-qubit patches can be initialized in $\ket{+} \otimes \ket{+}$ or $\ket{0} \otimes \ket{0}$, all with zero cost (0\clock). \textit{(b) Patch Deformation.} A patch can be expanded to cover additional tiles (1\clock) or shrunk to occupy fewer tiles (0\clock). 
By combining expansion and shrinkage, a patch can be moved to an adjacent tile (Fig.~\ref{fig:rules_cost}b). 
\textit{(c) Patch Rotation.} A patch can be rotated by combining corner movements and patch translation operations (Fig.~\ref{fig:rules_cost}c). \textit{(d) Measurement.} The product of Pauli operators can be measured when the relevant edges are adjacent to the \textit{ancilla path} or \textit{routing space}. This operation incurs a cost of 1\clock\ (Fig.~\ref{fig:rules_cost}d). Multi-patch $\pi/4$ and $\pi/8$ measurements can be performed by initializing an ancilla patch A, following the protocol proposed in~\cite{Litinski2019}. A summary of these rules is provided in Fig.~\ref{fig:rules_cost}. For detailed implementation of protocols, we refer to~\cite{Litinski2019}.

\textbf{Representative Data Layouts.} \textit{(a) Compact layouts}~\cite{Litinski2019} place qubit patches sequentially within a limited number of rows (e.g., 4 rows and columns in Fig.~\ref{fig:many_obv}). As more qubits are needed for later applications, additional columns are added to extend the layout, placing new patches (e.g.,  $q_4$, $q_5$) adjacent to existing ones. \textit{(b) Sparse layouts}~\cite{hirano2025localityaware} place both the $X$- and 
$Z$-edges of each data patch adjacent to the routing space, ensuring direct accessibility for logical operations. Each data patch is separated by at least one empty tile from its neighbors. As illustrated in Fig.~\ref{fig:many_obv}, this results in a sparse layout on a 4$\times$4 board. \textit{(c) Standard layouts}~\cite{hirano2025localityaware} are similar to sparse layouts, but use a different placement, as shown in Fig.~\ref{fig:many_obv}.  

\subsection{Pipeline of Executing Logical Circuits}

We introduce the top-down compilation flow following the practices outlined in~\cite{beverland2022assessing, hirano2025localityaware}. The pipeline is shown in Fig.~\ref{fig:LS_pipeline}.

\textbf{Step \circled{1}: Clifford+$T$ Decomposition.} The program begins with a quantum algorithm written in a high level-language. Since such algorithms are not expressed in the Clifford+$T$ gate set required for fault-tolerant execution, gate synthesis becomes necessary. This is typically done using the Solovay-Kitaev algorithm~\cite{dawson2005solovaykitaevalgorithm} or more advanced techniques~\cite{ross2016optimalancillafreecliffordtapproximation}.

\textbf{Step \circled{2}: Transpilation to Pauli-Based Computation.} The decomposed Clifford+$T$ circuits are subsequently transpiled into Pauli product rotations. The Pauli product rotations are defined as $P_{\theta} = \exp(-iP\theta)$, where $P$ is a multi-qubit Pauli operator. It is equivalent that $S = Z_{\pi/4}$ and $T = Z_{\pi/8}$ and the standard decompositions are given as: $H = Z_{\pi/4}X_{\pi/4}Z_{\pi/4}$ and $CNOT = (Z\otimes X)_{\pi/4} (I\otimes X)_{-\pi/4} (Z\otimes I)_{-\pi/4}$. There are several rules for simplifying circuits based on the commutation relations of Pauli operators.
If $P$ and $P'$ commute i.e. $PP' - P'P = 0$, then $P_{\pi/4}$ can be moved past $P'_{\theta}$. If $P$ and $P'$ anti-commute i.e. $PP' + P'P = 0$, $P'_{\theta}$ turns into $(iPP')_{\theta}$ when passing $P_{\pi/4}$. 
Clifford gates can be commuted through the circuit and absorbed into final measurements, as they map Pauli operators to Pauli operators.

\textbf{Step \circled{3}: Surface Code Level Mapping and Scheduling.} After transpilation of Pauli product rotations, one need to perform these instructions by mapping and scheduling following the rules required by lattice surgery. The instructions are initially mapped by assigning logical qubits to different patches of data layout, with the goal of maximizing opportunities for simultaneous multi-patch measurements while minimizing time costs. These instructions are executed sequentially according to the rules outlined in Fig.~\ref{fig:rules_cost}.

\subsection{Motivation}

\begin{figure}[h]
    \centering
    \includegraphics[width=1\linewidth]{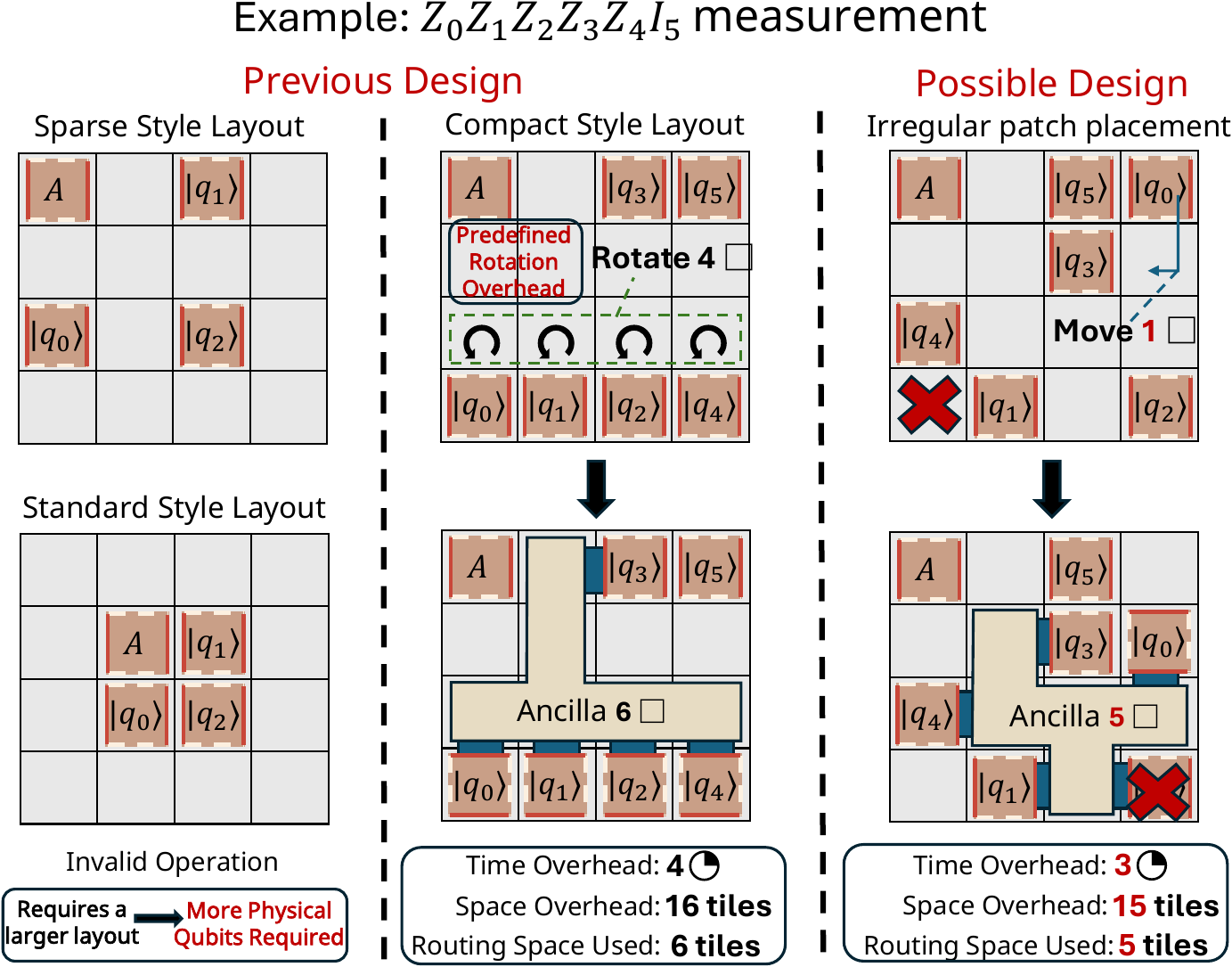}
    \caption{Example of layout design and loose scheduling (instruction rules based on Fig.~\ref{fig:rules_cost}). Prior sparse or standard style layouts (left) often suffer from inefficient resource usage, while compact style layout (middle) can incur additional scheduling overhead. In this context, irregular designs (right) can achieve higher patch utilization and fewer time steps.
    }
    \label{fig:many_obv}
\end{figure}

\begin{figure*}[t]
    \centering
    \includegraphics[width=1\linewidth]{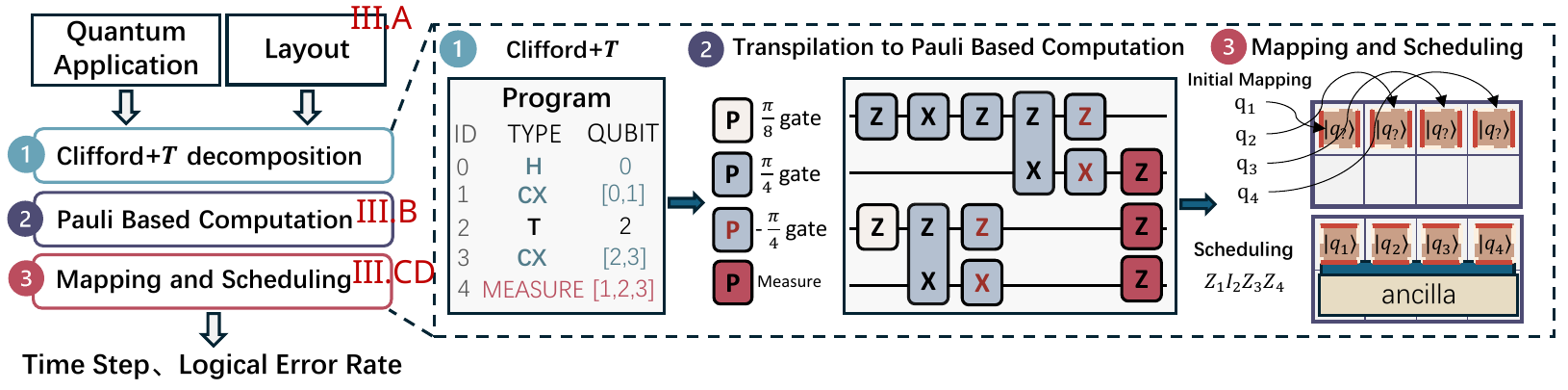}
    \caption{Pipeline of executing logical circuits. In this work, we introduce an algorithm to optimize layout design, equipped with advanced synthesis techniques for Pauli-based computation, as well as mapping and scheduling strategies to improve both time steps and logical error rate.}
    \label{fig:LS_pipeline}
\end{figure*}

\textbf{Observation 1 - Existing methods rely on fixed data layout.} 
In most cases, the layout is predefined or designed for executing logical operations.
However, these layouts may overlook the scheduling potential based on the specific shapes of patches and often require increasing the overall size to support larger quantum applications. For example, in Fig.~\ref{fig:many_obv}, consider performing a multi-patch measurement of $Z_0 Z_1 Z_2 Z_3 Z_4$ using the compact-style layout, where `A' denotes the ancilla patch reserved for potential operations. This operation requires 4 time steps due to patch rotation overhead and utilizes 6 ancilla patches to perform the measurements. In addition, the sparse-style layout is invalid in this case because it places an insufficient number of data qubit patches on the board, limiting the available logical resources for computation. 

In contrast, irregular data patch placement can potentially improve execution time and reduce space overhead, as indicated by the red cross. Additionally, since the data qubit patches are placed closer together, fewer ancilla patches are needed for routing. For example, Fig.~\ref{fig:many_obv} (right) requires only 5 ancilla routing patches. These potential benefits motivate the development of an automated search framework for designing layouts that are difficult to optimize manually.

\textbf{Observation 2 - Current compiler passes' scheduling are static.}
The term `static' refers to the current compiler passes that uniformly apply fixed scheduling strategies, regardless of circuit context or layout constraints. 
For example, in the compact layout, patches are typically rotated to align $X$ and $Z$ operators for multi-patch measurements. However, this approach overlooks cases where such in-place rotations are unnecessary. A concrete example is shown in Fig.~\ref{fig:many_obv} (right), where only $q_0$ is moved downward and the patch is rotated to expose a different edge to the routing space. This avoids the overhead of a full patch rotation and thereby reducing the required time steps. We refer to this more flexible strategy as `loose' scheduling, as it adapts the placement and orientation of logical qubit patches based on the circuit requirements.

\begin{figure}[h]
    \centering
    \includegraphics[width=1\linewidth]{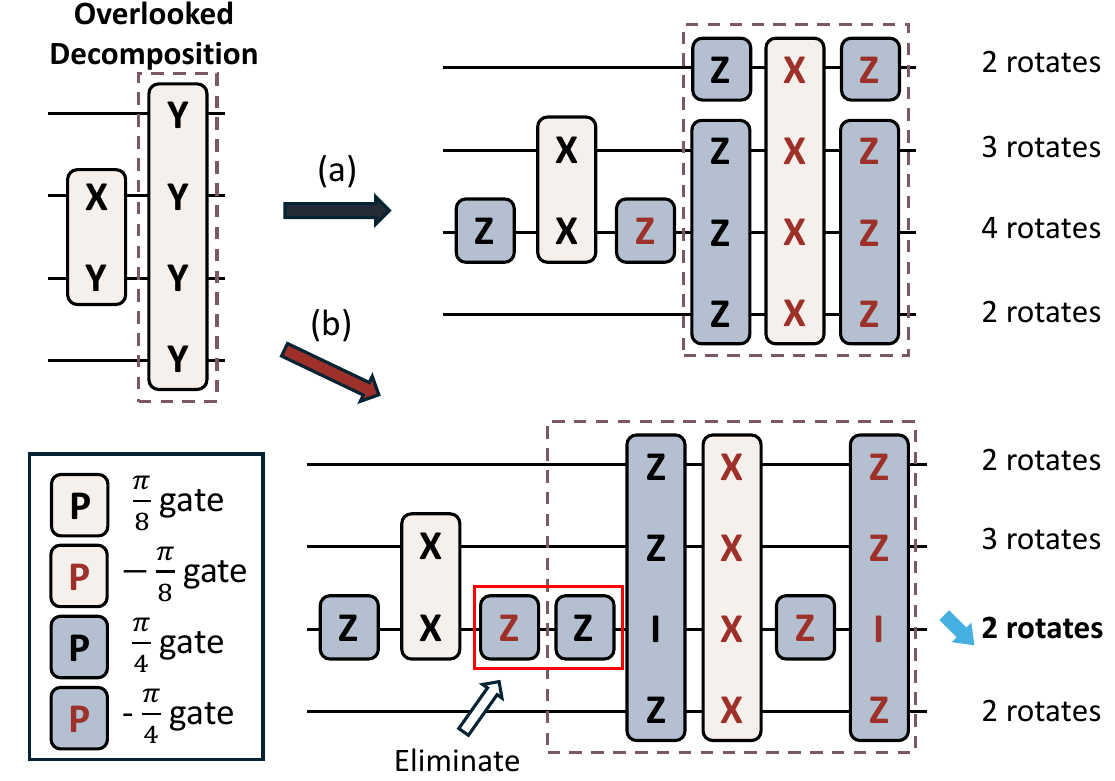}
    \caption{Operator cancellation opportunities in the process of Pauli-Y decomposition. (a) shows the decomposition method used in prior compiler passes~\cite{watkins2024high, leblond2023realistic}, which overlooks the cancellation opportunities shown in (b). 
    }
    \label{fig:y_synthesis_obv}
\end{figure}

\textbf{Observation 3 - Potential for Pauli operator cancellation.} 
In a more compact data layout where X and Z operators may not be accessible simultaneously, decomposition of the Y operator becomes necessary. An odd number of Y gates can be decomposed into a single rotation, while an even number must be decomposed into two. The latter enables multiple decomposition schemes, which can be exploited for potential gate cancellation. We present an example in Fig.~\ref{fig:y_synthesis_obv} that illustrates the synthesis of $Y$ operators using $X$ or $Z$ operators. In Fig.~\ref{fig:y_synthesis_obv}(a), the previous compiler pass~\cite{watkins2024high, leblond2023realistic} decomposes the rotations $(Y^{\otimes N})_{\pi/8}$ as $[Z_{\pi/4}\otimes (Z^{\otimes N-1})_{\pi/4}]  (X^{\otimes N})_{-\pi/8} [Z_{-\pi/4}\otimes (Z^{\otimes N-1})_{-\pi/4}]$ for even $N$, a method that overlooks potential cancellations among Pauli operators. In contrast, Fig.~\ref{fig:y_synthesis_obv}(b) shows a case where such cancellations can occur and be utilized in the synthesis process. We instead decomposing $(Y^{\otimes N})_{\pi/8}$ as $[(Z^{\otimes n})_{\pi/4}\otimes (Z^{\otimes N-n})_{\pi/4}]  (X^{\otimes N})_{-\pi/8} [(Z^{\otimes n})_{-\pi/4}\otimes (Z^{\otimes N-n})_{-\pi/4}]$ where $n$ is odd then $N-n$ is odd. This enables operator elimination and offers potential to reduce time costs.

\begin{figure}[h]
    \centering
    \includegraphics[width=1\linewidth]{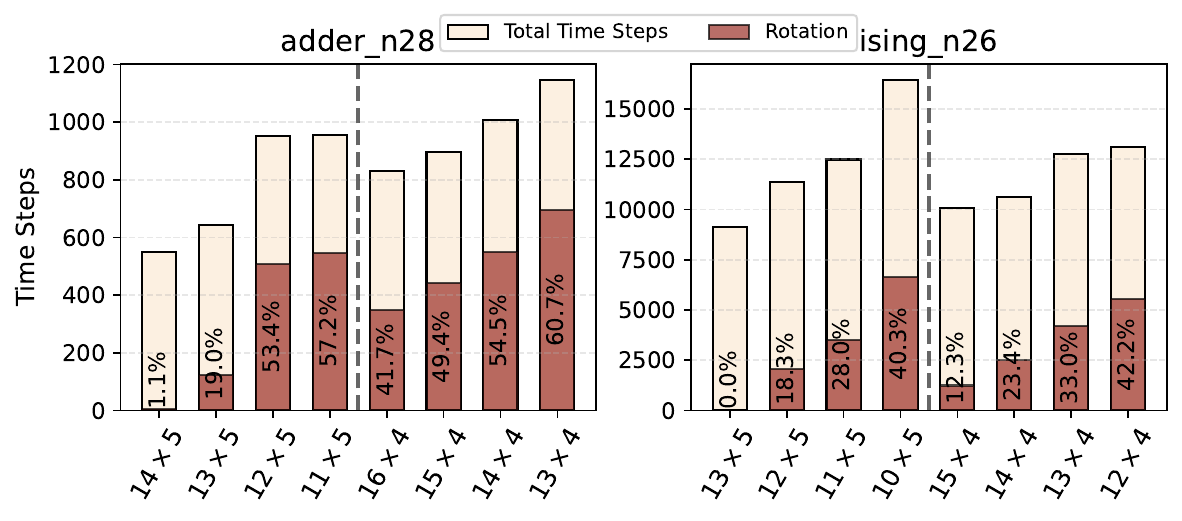}
    \caption{Quantification of rotation bottleneck (data-layout sizes in x-axis).}
    \label{fig:rotation_bottleneck}
\end{figure}

\textbf{Observation 4 - Rotations are the primary bottleneck in space-constrained layouts.} In space-constrained or irregular data layouts, such as the one shown in Fig.~\ref{fig:many_obv}, the primary factor limiting execution speed is the need to expose the $Z$ operator to the routing space from the $X$ operator, which requires a rotation operation that takes 3 time steps. This issue is especially evident in irregular layouts, where patches differ in their access to X and Z operators. Some patches expose only a single X or Z operator to the routing space, while others expose both. If a qubit that frequently switches between $X$ and $Z$ operators is mapped to a patch that supports only one of them, the resulting overhead can be significant. This overhead is quantified in Fig.~\ref{fig:rotation_bottleneck}, we observe that as the total number of tiles decreases, rotations could become the dominant bottleneck, accounting for more than 50\% of the overall time steps. In contrast, using an optimized layout size with moderate tile count, the rotation overhead can be reduced. This motivates us to design improved data layouts that balance tile count and rotation overhead. 

\section{The O3LS Compiler}\label{sec:compiler_method}

\subsection{O3LS Module 1: Layout Design} \label{sec:layout_design}

We first propose an algorithm for the automated design of logical qubit data layouts, enabling the search for more squeezed configurations. The core design principle is to preserve the connectivity between the routing space and all data patches, while also maximizing the number of $X$ and $Z$ edges that each data patch exposes to the routing space. This is because the connectivity is intentionally designed to ensure that all data patches are measurable, thereby preventing the failure of logical operations. Moreover, the primary source of time overhead arises from patch rotations. If the layout allows for a greater variety of edge types, this overhead can be reduced, potentially improving overall execution efficiency.

\textbf{Layout Design Scoring Function.} Based on these intuitions, we first design a scoring function $S$ to evaluate the goodness for a given board $B$. The detailed formulation of the scoring function $S$ is as follows:
\begin{equation}
    S(B)=C(B)\times(N_x(B)+N_z(B)-\alpha_e N_e(B)).
\end{equation}
We use \( C(B) \) to indicate whether a routing path exists in the given board or designed layout \( B \) that connects to at least one edge of specified data patches. We further require that both the $x$- and $z$-edges of the ancilla patch be connected to the routing space, as completing Pauli product rotations necessitates performing $Y$ measurements on the newly initialized ancilla patch~\cite{Litinski2019}. The value of \( C(B) \) is 1 if such a routing path exists, and 0 otherwise.  We use \( N_x(B) \) to denote the number of qubits on board \( B \) that are connected to the \( x \)-edge by a routing path. Similarly, \( N_z(B) \) represents the number of qubits connected to the \( z \)-edge. 
Additionally, we introduce \( N_e(B) \) as a penalty term, defined as the number of edges in \( B \) that connect data patches to the routing space, and is penalized by the density factor $\alpha_e$.
This term primarily guides the layout design by encouraging the subsequent design process to favor either more compact or sparser layouts.

\begin{figure}[t]
    \centering
    \includegraphics[width=1.0\linewidth]{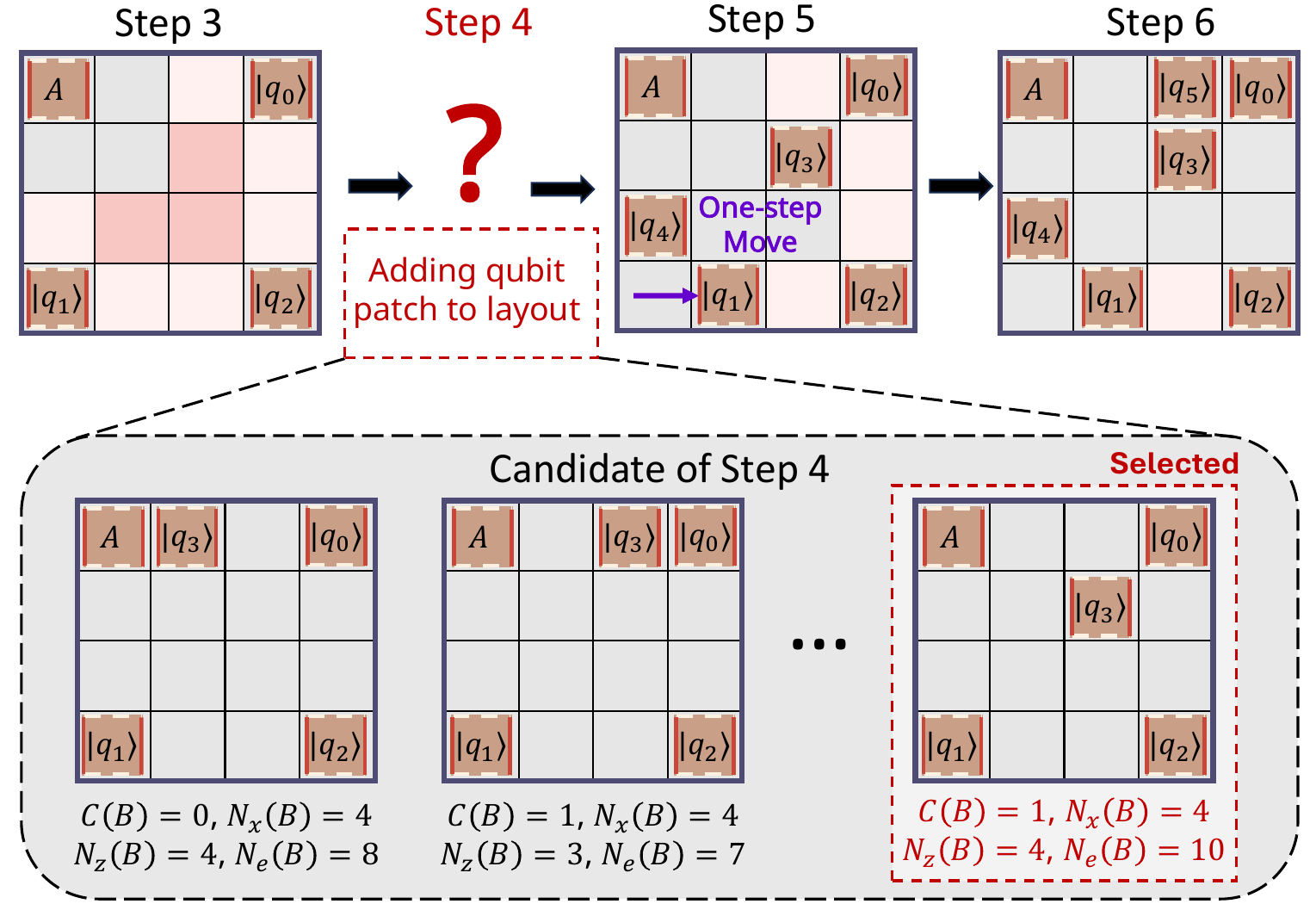}
    \caption{Layout design process. Darker pink color indicate higher scores. At each step, the highest-scoring position is selected as the new qubit patch.}
    \label{fig:layout_design_example}
\end{figure}
\textbf{Design Process.} Based on the scoring function, we propose an iterative layout design method. The board is initialized with an ancilla patch A placed at the corner. Then, at each step, we attempt to add a data patch to the current board \( B \), generating a list of candidate boards \(\{B_1^{(1)}, B_2^{(1)}, \cdots\}\). Among these candidates, the board with highest-scoring \( B_i^{(1)} \) is selected as the updated layout. This process is repeated iteratively until all data patches have been successfully placed. Fig.~\ref{fig:layout_design_example} shows this procedure, where positions indicating the routing path are marked in gray. At each iteration, the location with the highest score is selected for placing the next data patch.

Additionally, after each new qubit patch is placed, we perform a post-processing step to further enhance the overall layout performance. Specifically, this step involves evaluating the potential for relocating existing qubit patches through a one-step move within the current board configuration. 
If a modified board \( B' \) resulting from such a relocation yields a higher score according to the scoring function, then \( B' \) is adopted as the updated layout. 
An example of this process is shown in Step 5 of Fig.~\ref{fig:layout_design_example}. After placing \( q_4 \), relocating \( q_5 \) increases the number of distinct edges exhibited by \( q_5 \) while preserving routing connectivity. This one-step adjustment serves as an effective means of improving the layout’s overall performance. 

\textbf{Complexity Analysis.}
Overall, the computational complexity is $\mathcal{O}(n|B|)$, where $n$ denotes the number of qubits and $|B|$ represents the size of board.

\subsection{O3LS Module 2: Y-Synthesis Algorithm}\label{sec:module1}

We then introduce O3LS module 2 to address observation 3: Y-operator decomposition is necessary, and previous compilers have missed opportunities for Pauli operator cancellation in squeezed data layouts. The key motivation stems from the fact that measuring a Pauli-$Y$ operator requires simultaneous access to both $X$ and $Z$ operators. However, in many layouts with limited data patches, such simultaneous measurements are not feasible, including these generated from O3LS Module 1. This limitation necessitates decomposing Pauli Y operators into equivalent combinations of Pauli X and Z operators.

\begin{algorithm}[t]
    \renewcommand{\algorithmicrequire}{\textbf{Input:}}
    \renewcommand{\algorithmicensure}{\textbf{Output:}}
    \begin{algorithmic}[1]
        \REQUIRE Pauli operator sequence $S = \{P_1, P_2, \dots, P_l\}$
        \ENSURE Synthesized Pauli operator sequence $S'$
        \STATE Initialize $S'=\{\}$
        \FOR{$P_i\in S$}
        \STATE Get the Y indices of $P_i$ as $\text{Y\_indices}_i$
        \IF{size($\text{Y\_indices}_i$) == 0}
        \STATE $S'.append(P_i)$
        \STATE Continue
        \ELSIF{size($\text{Y\_indices}_i$) is odd}
        \STATE 
        Set $b_1=\text{Y\_indices}_i$ and $b_2=\varnothing  $
        
        \ELSE
        \STATE Find the bipartition $b_1, b_2$ of the set \( \text{Y\_indices}_i \).
        \ENDIF
        \STATE Build left Z-rotation operator $L_i^{(1)},L_i^{(2)}$ and right Z-rotation operator $R_i^{(1)},R_i^{(2)}$ according to $b_1, b_2$
        \STATE Decompose $P_i$ to get Y-free operator $P'_i$
        \STATE $S'.append(L_i^{(1)},L_i^{(2)},P'_i, R_i^{(1)},R_i^{(2)})$
        \ENDFOR
        \STATE Do Pauli operator synthesis on $S'$
        \RETURN $S'$
    \end{algorithmic}
\caption{Y-synthesis Algorithm}
\label{algo:y synthesis}
\end{algorithm}

\begin{figure}[h]
    \centering
    \includegraphics[width=1\linewidth]{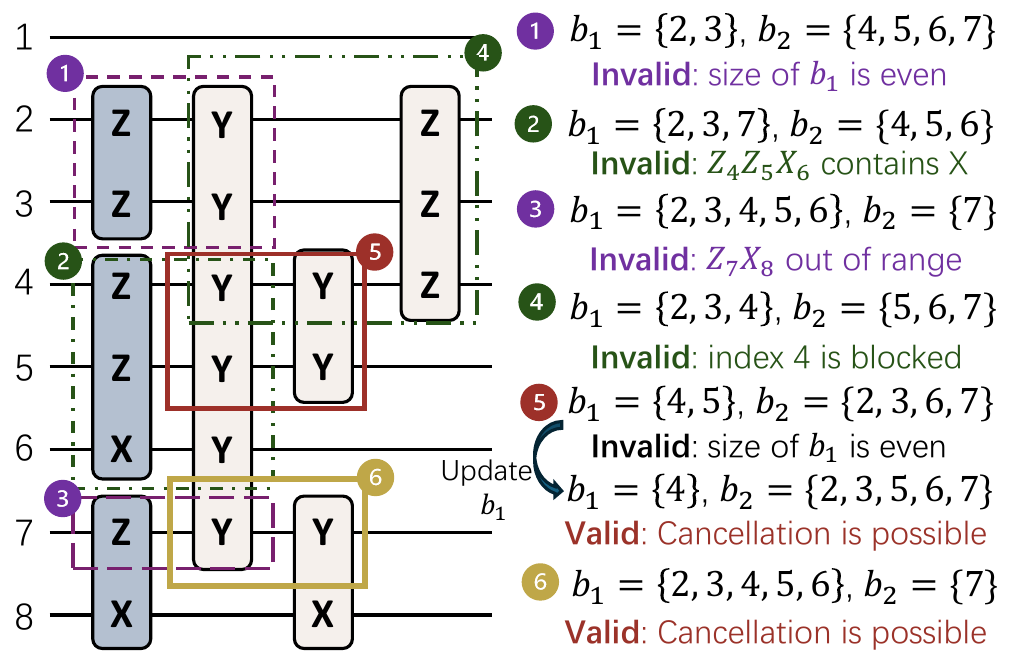}
    \caption{An example of Y-synthesis rule.
    }
    \label{fig:y_decompose_rule}
\end{figure}

\textbf{Y-Synthesis Algorithm.} The Y-synthesis algorithm is structured as a two-step process.
\circled{1} \textit{Y-decompose.} The process referred to as Y-decomposition (e.g. step 3-13 in Algorithm~\ref{algo:y synthesis}) involves transforming Pauli-$Y$ operators into combinations of Pauli-$X$ and Pauli-$Z$ operators, since direct $Y$-basis measurements are not always supported by the available patch configurations on the surface code lattice. As illustrated in Fig.~\ref{fig:y_decompose_rule}, there may exist multiple valid decomposition schemes for the same operator. Therefore, the Y-decomposition process must select the decomposition that is most suitable for Pauli operator synthesis, particularly in terms of enabling Pauli operator cancellation. 
\circled{2} \textit{Pauli Operator Synthesis.} Pauli operator synthesis (e.g. step 16 in Algorithm~\ref{algo:y synthesis}) is the process of taking a sequence of Pauli operators and merging adjacent ones that share the same Pauli basis to enable cancellation and reduce circuit overhead.

The Y-Synthesis pseudocode is given in Algorithm~\ref{algo:y synthesis}, whose Step 10 is pivotal.
Due to the inherent properties of Pauli operators, if an operator contains an even number of Y components, these Y components must be divided into two non-overlapping groups, each containing an odd number of Y components. Subsequently, left Z-rotation and right Z-rotation operators are constructed based on these groups. 

During this process, Y-synthesis algorithm evaluates whether any operator derived from a group can be absorbed by other operators. Specifically, for each group, we check if at least one operator within that group can be absorbed by another operator. If such an operator exists, the group is designated as a candidate. Among these candidate groups, we then select the one with the highest number of absorbed operators. 
Fig. \ref{fig:y_decompose_rule} is an example of the algorithm, which illustrates the relevant constraints. 
Since the input Pauli operator sequence is a topologically sorted list, for a specific Pauli operator \( P_i \), its predecessor operators must have already completed the Y-decompose process. Therefore, it is sufficient to ascertain whether the group can be absorbed by its predecessor operators. 
Additionally, for the successor operators, it is necessary to verify whether there exists a potential Y-decomposition opportunity that could enable the group to be absorbed.

\textbf{O3LS-IR.}
To support efficient Y-synthesis in O3LS, we introduce O3LS-IR, an intermediate representation that captures dependencies among Pauli operations to determine execution order and enable parallelism. Pauli operators are represented as nodes in a Pauli Directed Acyclic Graph (PDAG), recorded in our quantum IR, which includes: 
\begin{enumerate}
\item \textit{Rotation angle:} This specifies the rotation angle associated with the Pauli operator. The rotation angle can also indicate whether the node represents a measurement. 
\item \textit{Pauli words:} These represent the Pauli operators.
\item \textit{Predecessor nodes:} These are nodes that precede the current node in the DAG.
\item \textit{Successor nodes:} These are nodes that follow the current node in the DAG.
\end{enumerate}

The rules for constructing the PDAG are as follows: Each Pauli operator is represented as a node in the PDAG. A directed edge \((P_i, P_j)\) exists between two nodes \(P_i\) and \(P_j\) if and only if:
(1). There is at least one qubit \(q\) on which both \(P_i\) and \(P_j\) are not the identity operator \(I\).
(2). No other Pauli operator \( P_k \) between \( P_i \) and \( P_j \) acts non-trivially on qubit \( q \). A node with in-degree 0 corresponds to an executable Pauli operator. Once executed, the node is removed, and the process repeats until all nodes are processed, ensuring correct dependency resolution.

\textbf{Complexity analysis.} 
During Y-synthesis, each operator examines at most $n$ predecessor and $n$ successor nodes, resulting in a per-operator computational complexity $\mathcal{O}(n)$. Considering that the Y-decomposition involves at most $l$ Pauli operators, the overall complexity of Algorithm~\ref{algo:y synthesis} is $ \mathcal{O}(nl) $.

\subsection{O3LS Module 3: Loose Scheduling}\label{sec:module2}

As highlighted in Observation 2 and illustrated in Fig.~\ref{fig:many_obv}, most existing approaches rely on pre-defined scheduling patterns or assume that logical patches are sufficiently large to avoid additional patch rotations during multi-patch measurements. 
This leaves room for further optimization in scheduling. To address this, the O3LS module implements a loose scheduling strategy, where loose refers to the flexibility of the scheduling process. 
This approach allows for dynamic repositioning and adjustment of patches to better accommodate measurement requirements, ultimately aiming to reduce the total number of time steps. The pseudocode for the loose scheduling algorithm is provided in Algorithm~\ref{algo:dynamic scheudlie}.

\begin{algorithm}
    \renewcommand{\algorithmicrequire}{\textbf{Input:}}
    \renewcommand{\algorithmicensure}{\textbf{Output:}}
    \begin{algorithmic}[1]
        \REQUIRE Pauli operator sequence $S = \{P_1, P_2, \dots, P_l\}$, qubit number $n$, Board $B$
        \ENSURE Executable operation sequence $S'$
        \STATE Initialize $S'=\{\,\}$
        \STATE Build Pauli DAG $\mathcal{G}$ from  $S$
        \WHILE{$\mathcal{G}$ is not empty}
        \FOR{$P_i\in\mathcal{G}$.frontier and $P_i$ is executable }
        \STATE Get bus patch list $L_{p_i}$ for $P_i$
        \STATE Execute $P_i$ on $L_{p_i}$ and update $S'\leftarrow S'+P_i$
        \ENDFOR
        \STATE Pop a Pauli operator $P_j$ from $\mathcal{G}$.frontier 
        \WHILE{$P_j$ is not executable}
        \STATE Get all possible patch operations $O_B$ from $B$
        \STATE Select \( o_b \in O_B \) with the best reward $r(o_b, P_j)$
        \STATE Execute $o_b$ on B and update $S'\leftarrow S'+o_b$
        \ENDWHILE
        \ENDWHILE
        \RETURN $S'$
    \end{algorithmic}
\caption{Loose Scheduling Algorithm}
\label{algo:dynamic scheudlie}
\end{algorithm}

\textbf{Determining bus patch.}
In Step 5, the objective is to identify the minimal amount of routing space required to execute \( P_i \), as reducing the overall routing path length not only increases opportunities for parallel execution of other operations but also has the potential to lower the overall logical error rate.
To achieve this, we apply Dijkstra's algorithm to sequentially determine the shortest paths between the required patches. Previously identified paths are treated as nodes with zero cost. This process results in a bus patch list with a minimal number of patches used and the complexity of each executed Pauli operator $P_i$ is $\mathcal{O}(|B|^2)$, where $|B|$ denotes the size of board.

\textbf{Resolving unexecutable Pauli operator.} In Step 11, the reward function is decomposed into three components. First, it is defined as the number of data patches in \( B_o \) that enable the execution of \( P_j \); that is the number of data patches for which a valid path exists to support the application of \( P_j \), where \( B_o \) denotes the state of the board \( B \) after applying the patch operation \( o \). 
Second, maintaining connectivity among all data patches is essential for enabling subsequent lattice surgery operations, so any patch operation that breaks this connectivity receives a reward of zero. 
Third, when multiple patch operations yield the same reward, preference is given to those incur lower time overhead.

\textbf{Complexity Analysis.}
Since the number of candidate patch operations for each data patch is constant, the total number of candidates is $\mathcal{O}(n)$, where $n$ denotes the number of data patches. Also, the complexity of evaluating the reward function is also $\mathcal{O}(n)$. Furthermore, the reward function is designed to guarantee that each patch operation increases the number of patches satisfying the execution requirements by at least one. Consequently, the overall complexity of a complete scheduling process is $\mathcal{O}(n^2)$.

\subsection{O3LS Module 4: Initial Mapping}\label{sec:module3}

\textbf{Edge-aware Initial Mapping.} Among all patch operations, qubit patch rotations are a major source of time overhead, particularly in squeeze layouts where such operations occur more frequently. To mitigate this overhead, we propose an edge-aware (EA) initial mapping strategy. The core idea is to analyze the PDAG to estimate the rotation demand for each qubit. Qubits with higher expected rotation frequencies are preferentially mapped to patches that are adjacent to both the X and Z-edges of the ancilla patch, thereby minimizing the need for costly patch rotations.

The effectiveness of the EA mapping method is most evident in layout boards that fall between compact and sparse configurations. For instance, in highly compact layouts, changes in qubit positions introduced by the mapping are less likely to expose additional edges, thereby limiting the potential benefits of the approach. In these cases, the performance of the EA strategy becomes highly dependent on the specific structure and gate distribution of the input circuit. Moreover, in sparse qubit patch layouts, the edge-aware component contributes less to performance improvement. This is because, in sparse configurations, each patch can typically expose both X and Z operators simultaneously, reducing the advantage gained from edge-aware placement.

\textbf{Complexity Analysis.} The rotation demand required for EA mapping can be efficiently extracted during the construction of PDAG. The process of counting the number of edges exposed by data patches has a computational complexity of $\mathcal{O}(n)$, where $n$ denotes the number of data patches. The final mapping is achieved through two applications of quicksort, resulting in an overall complexity of \(\mathcal{O}(n \log n)\).

\begin{figure*}[t]
    \centering
    \includegraphics[width=1\linewidth]{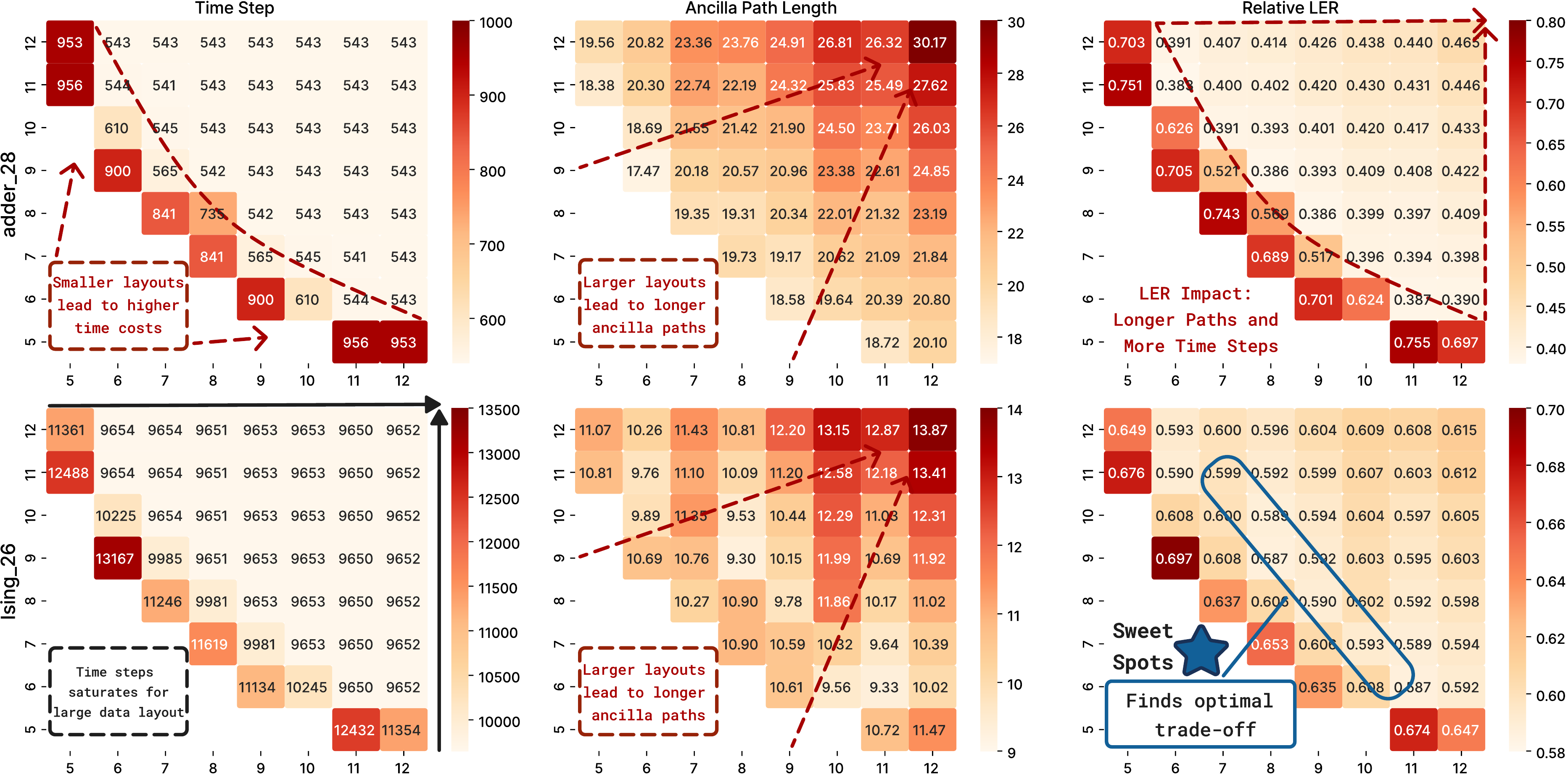}
    \caption{Performance of executing applications on various data layout sizes using O3LS, including their corresponding time steps, ancilla patch lengths, and logical error rates for adder\_28 and ising\_26. The X and Y axes represent patch boards of size $N\times M$, where $N$ and $M$ range from 5 to 12.
    }
    \label{fig:heat_map}
\end{figure*}

\section{Evaluation}

\subsection{Experiment Setup}
\textbf{ a) Metrics. } (1). \textit{Logical error rate.} We simulate the logical error rate (LER) by parsing lattice surgery instructions and analyzing them at the time-slice level. We then compute the layer-wise LER for each time slice and linearly accumulate them over the entire execution:
\begin{equation*}
\begin{aligned}
    p_{\text {total}}\approx\sum_{t=1}^T p_{\text {layer}}^{(t)}
    \approx \sum_{t=1}^T (1-(1 - P_{\text{PPM}}^{(t)})(1 - P_{\text{PR}}^{(t)})(1 - P_{\text{idle}}^{(t)})),
\end{aligned}
\end{equation*}
which follows the same method used in SPARO~\cite{kan2025sparo} under the assumption of rare failures and independent error events. Each $p_{\text{layer}}^{(t)}$ is estimated from the simulated LER in Pauli product measurement (PPM), patch rotation (PR), and idling memory errors during that layer. The PPM error rate is mainly determined by the routing space and the code distance. The rotation step is decomposed into three slices: patch deformation, corner movement, and patch movement~\cite{Litinski2019}, and they are simulated separately. (2). \textit{Time cost.} Following prior work~\cite{hirano2025localityaware, wang2024optimizing}, our evaluation also focuses on the time cost (\clock). We also record the size of data layout, which serves as an indicator of the space-time volume~\cite{Tan2024LaS}.

\textbf{b) Baseline Algorithms. } We evaluate O3LS against the \texttt{SPC} algorithm~\cite{Litinski2019}, as implemented in~\cite{watkins2024high, leblond2023realistic}, and the recent locality-aware method \texttt{LAPBC}~\cite{hirano2025localityaware}, which enhances circuit parallelism and outperforms prior compilers~\cite{beverland2022surface, molavi2025dependency}. We also compare against \texttt{SPARO}~\cite{kan2025sparo}, another automated data-layout design method that aims to expand data layouts.

\textbf{c) Benchmarks. }
We benchmark using a representative set of FT quantum algorithms, following prior FTQC compiler studies~\cite{trochatos2025trace, wang2024optimizing, kan2025sparo, molavi2025dependency, maurya2024managing}. These include circuits for Hamiltonian simulation, Quantum Fourier Transform, key components of Shor's algorithm (e.g., adders and multipliers), and SWAP tests for quantum machine learning, \textit{many of which serve as building blocks for larger algorithms.}
We source the QASM files from MQT Bench~\cite{Quetschlich_2023} and FTCircuitBench~\cite{harkness2026ftcircuitbenchbenchmarksuitefaulttolerant}. Some FTCircuitBench circuits were originally taken from QASMBench~\cite{li2023qasmbench}. Unless otherwise specified in the circuit name, we assume a one-dimensional Hamiltonian.
We evaluate O3LS across different layouts, including the compact design from~\cite{Litinski2019} and standard layouts from~\cite{hirano2025localityaware}.

\textbf{d) Experimental Setting.} The benchmarks are decomposed into Clifford+$T$ circuits using \texttt{GridSynth}~\cite{qiskitgridsynth}, based on~\cite{ross2016optimalancillafreecliffordtapproximation} with a synthesis error tolerance of $10^{-5}$. STIM simulations~\cite{gidney2021stim} are conducted to characterize atomic lattice surgery operations using a $d=9$ surface code under a circuit-level depolarizing noise model with a physical error rate of $p=10^{-3}$. Each atomic operation is independently compiled into a STIM circuit and simulated using Monte Carlo sampling with no less than $10^6$ trials. Decoding is performed by PyMatching 2~\cite{higgott2025sparse}. For all experiments, we use a magic-state factory based on the 15-to-1 distillation protocol~\cite{magicdistillation1}. The factory is placed outside the designed layout, while ensuring at least one routing path connects it to the data region. The $\pi/4$ and $\pi/8$ Pauli-product measurements are implemented via standard gate teleportation protocol, following the implementation in~\cite{Litinski2019} (Fig.~7 and 11(b)).

\textbf{e) Simulation Device.} 
All simulations were performed on a device with an Intel Core i9-14900K 32-core processor and 188 GB of RAM using Python 3.10.

\subsection{Analysis of Data Layout Designs}

\textbf{a) Performance on Data Layout Designs.}
We begin by analyzing the performanc across different data layout sizes and their corresponding estimated logical error rates. In this scenario, we consider patch boards of size $N \times M$, where $N$ and $M$ range from 5 to 12, and evaluate the performance of \text{adder\_28} and \text{Ising\_26} circuits across various patch board configurations. The experimental results are presented in Fig.~\ref{fig:heat_map}, where the heatmaps illustrate the time steps, ancilla path lengths, and their associated RLER.

Our evaluation reveals that when the patch board size is too small, the primary performance bottleneck arises from the overhead associated with operation scheduling, such as patch rotations. As the board size increases, the number of time steps required for execution decreases and eventually converges. However, despite similar time step at larger board sizes, the ancilla patch length increases monotonically due to the availability of more routing space. This leads to an observation that both higher time step and longer ancilla path lengths contribute significantly to increased logical error rates. \textit{These results underscore a fundamental trade-off between time costs and ancilla patch distance, indicating that carefully designed, smaller data layouts can contribute to lower logical error rates.}

On the other hand, the applications that are required to run on these fixed data layouts necessitate the use of $10\times 10$ and $9\times15$ patch boards in the standard and sparse layouts. Compared to these scenarios, O3LS effectively reduces the required space overhead, achieving a board size reduction of up to 28.0\% and 46.7\%, while preserving the number of time steps. Furthermore, it achieves a reduction in logical error rates of up to 16.9\% compared to larger data layouts (e.g., $12\times12$), due to the decreased length of ancilla patches. While larger data layouts can reduce the number of time steps, they often require longer routing paths. In contrast, smaller data layouts are more space-efficient but tend to suffer from higher time costs. Both scenarios can contribute to increased logical error rates. To balance this trade-off, \textit{O3LS generates more compact layout designs that minimize ancilla patch length while maintaining time costs comparable to those of sparse layouts, thereby achieving a sweet spot for reducing the overall logical error rate.}

\begin{figure}[t]
    \centering
    \includegraphics[width=0.95\linewidth]{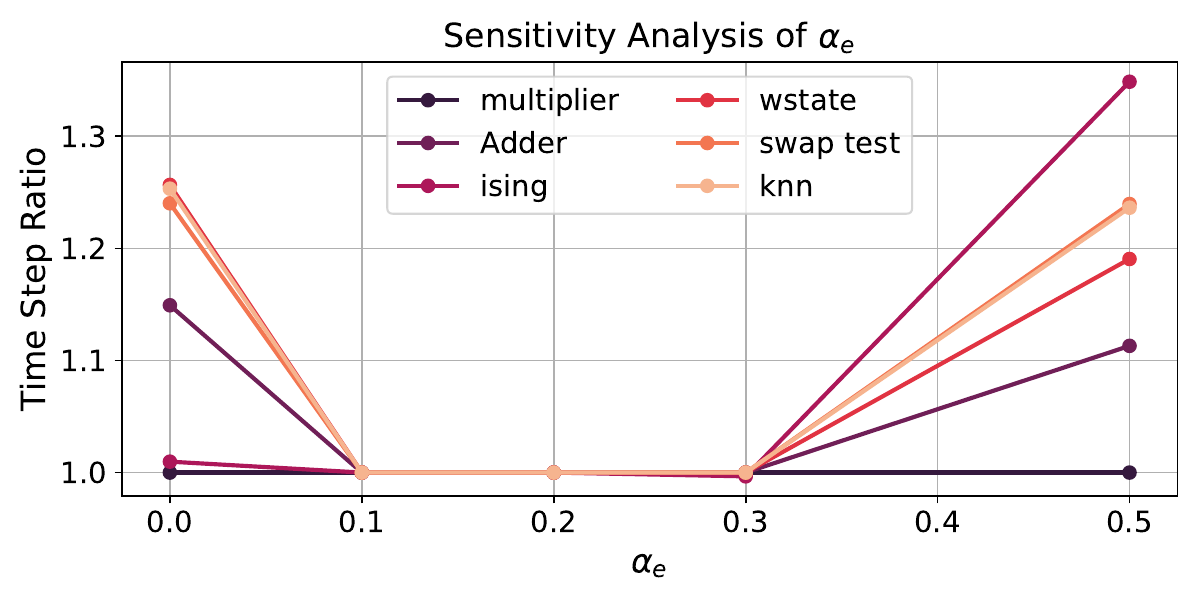}
    \caption{Sensitivity analysis of density factor on automatic layout design. The relative time step is defined as the ratio between the evaluated time step and the baseline time step corresponding to density factor \( \alpha_e = 0.1 \).}
    \label{fig:sensitivity_analysis}
\end{figure}

\textbf{b) Sensitivity Analysis on Density Factor.}
We also perform a sensitivity analysis of the density factor \(\alpha_e\) in the layout design, as introduced in Section~\ref{sec:layout_design}, with results presented in Fig.~\ref{fig:sensitivity_analysis}. The experiments indicate that values of \(\alpha_e\) between 0.1 and 0.3 yield the best performance across most applications. Additionally, performance remains relatively stable for $ 0 < \alpha_e < 0.5 $, but degrades noticeably when \( \alpha_e = 0 \) or \( \alpha_e = 0.5 \). This behavior can be attributed to the trade-off governed by $\alpha_e$: smaller values prioritize placing patches with multiple edges to ancilla patches, promoting sparse layouts, while larger values emphasize compactness. Both extremes can lead to suboptimal layouts—either overly sparse and causing later qubit patches to obscure certain operators or overly compact and underutilizing available patches. Thus, an appropriate balance of $\alpha_e$ is essential to optimize the trade-off between compact and sparse placement in patch utilization. We recommend using $\alpha_e \in [0.1, 0.3]$ to achieve this balance.

\begin{figure}[h]
    \centering
    \includegraphics[width=1\linewidth]{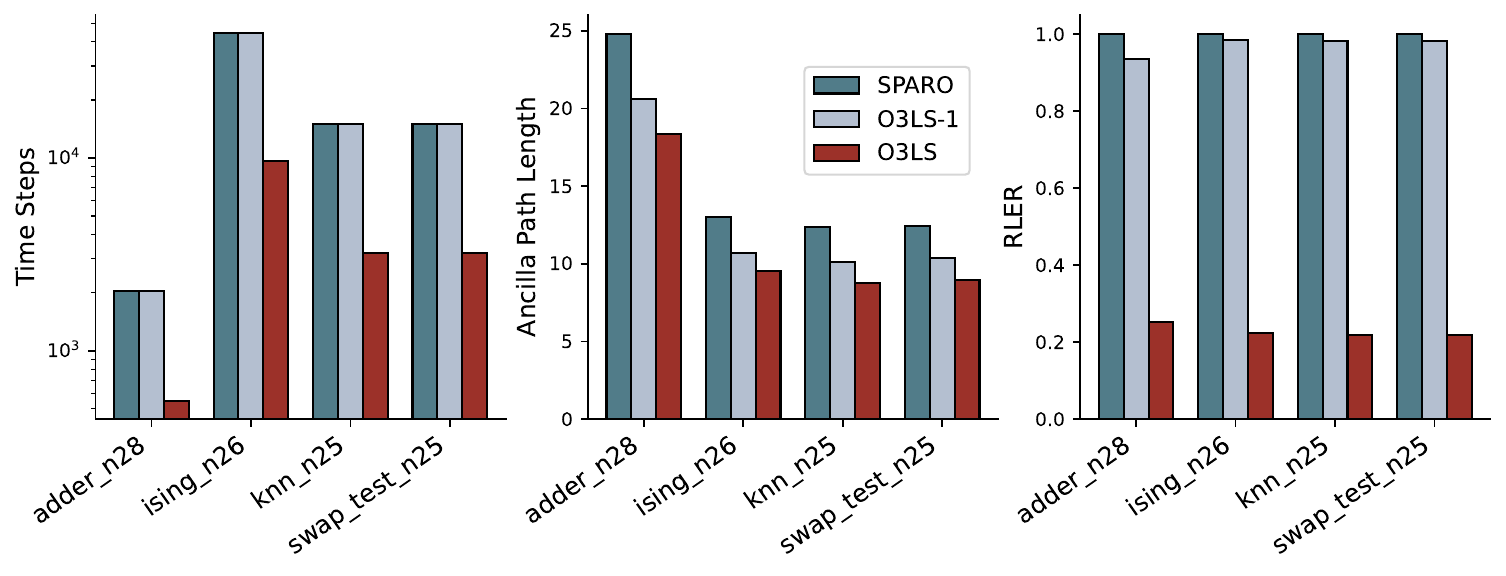}
    \caption{Layout design comparison with SPARO~\cite{kan2025sparo}.}
    \label{fig:layout_design_comparison}
\end{figure}

\textbf{c) Comparison with SPARO.} Moreover, other layout-design methods such as SPARO~\cite{kan2025sparo} explore strategies to improve data-layout utilization. In Fig.~\ref{fig:layout_design_comparison}, we compare O3LS with SPARO’s layout-design approach. `O3LS-1' uses an O3LS-generated data layout with SPARO’s scheduling method and `O3LS' uses the full O3LS stack for both layout generation and scheduling. 

O3LS-generated data layouts achieve better performance than SPARO, improving LER by 3.05\% on average when paired with SPARO's scheduling method. This gain comes primarily from smaller data layouts, which reduce average ancilla-routing space by 17.35\%. In addition, O3LS’s scheduler is particularly effective on the smaller data layouts produced by O3LS Module 1. It reduces time steps by 78.24\% and average routing space by 27.17\% on average, which together yield a further 77.1\% reduction in LER. 
The routing-space savings in O3LS largely stem from its objective. O3LS searches for a sweet spot that minimizes space overhead in the data layout, whereas SPARO tends to allocate more data-layout resources. O3LS also includes more advanced synthesis and a looser scheduling strategy, which further reduces time steps. Overall, O3LS finds smaller layouts that shorten ancilla paths while reducing time steps, resulting in lower LER.

\begin{figure*}[t]
    \centering
    \includegraphics[width=1\linewidth]{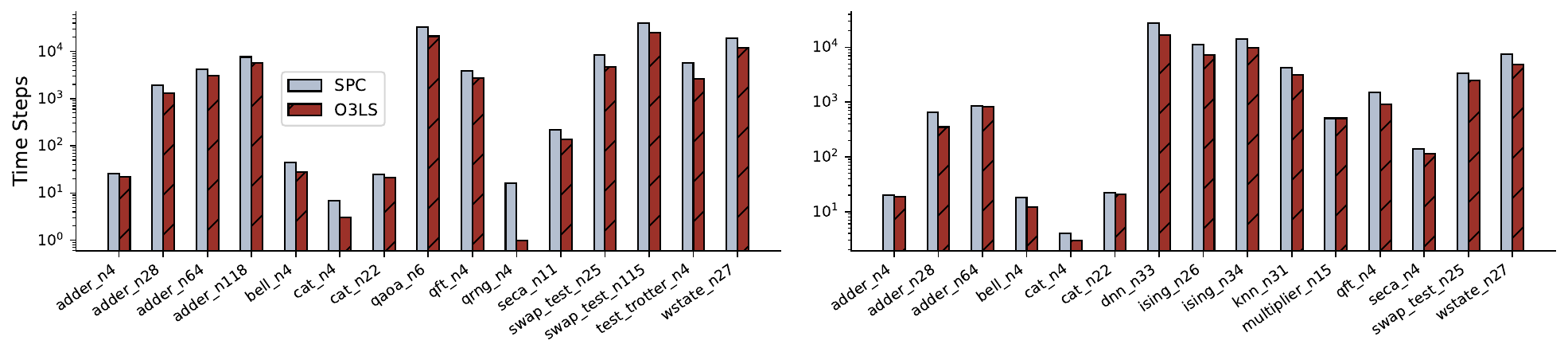}
    \caption{Compilation technique comparison with \texttt{SPC}. (Left) Results with the compact layout. (Right) Results with the standard layout. }
    \label{fig:direct_comparison}
\end{figure*}
\begin{figure*}[t]
    \centering
    \includegraphics[width=1\linewidth]{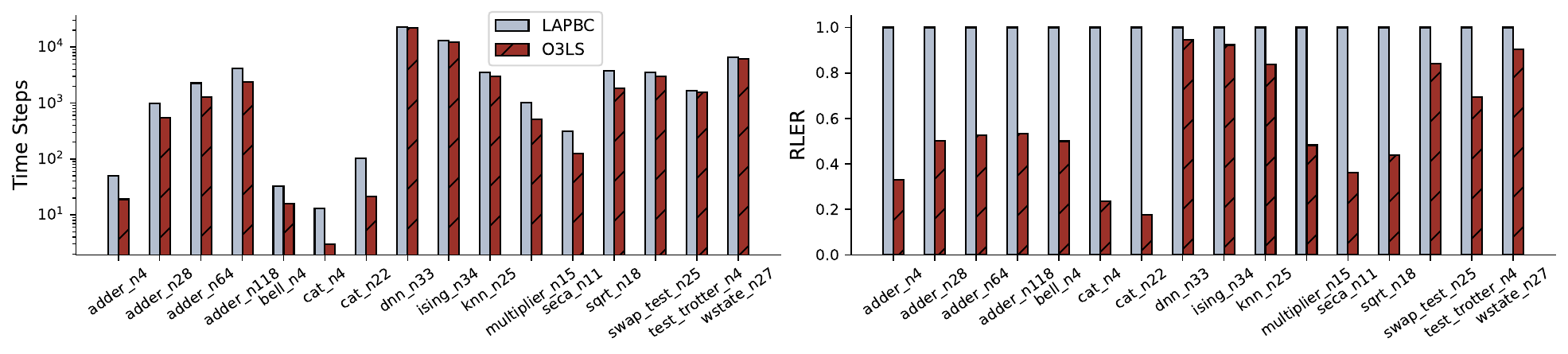}
    \caption{(Left) Comparison of time steps between the prior compiler pass \texttt{LAPBC}, which emphasizes parallelism, and our proposed O3LS. (Right) Logical error rate analysis indicates that O3LS achieves superior performance in most scenarios, highlighting the effectiveness of the proposed scheduling and Y-synthesis algorithms.
    }
    \label{fig:parallel_comparison}
\end{figure*}

\subsection{Compilation Technique Comparison}

\textbf{ a) Comparison with \texttt{SPC}.} We begin by comparing our compiler pass with \texttt{SPC} across two data layout configurations: compact and standard. 
As shown in Fig.~\ref{fig:direct_comparison}, our compiler pass O3LS achieves an average reduction of 36.07\% in time steps compared to \texttt{SPC} under the compact layout, and an average reduction of 24.76\% under the standard layout. These improvements achieved by O3LS are due to the fact that \texttt{SPC} does not incorporate any optimization techniques, such as scheduling, routing, or synthesis.

\textbf{ b) Comparison with \texttt{LAPBC}.} Recent efforts have focused on improving circuit parallelism in lattice surgery~\cite{beverland2022assessing, beverland2022surface, hamada2024latticesurgery, hirano2025localityaware}, with \texttt{LAPBC}~\cite{hirano2025localityaware} as the latest advancement.
In Fig.~\ref{fig:parallel_comparison} (left), we analyze the time steps required by both compilers. The results show that O3LS, achieves an average time-step reduction of 35.10\% compared to \texttt{LAPBC}, with a maximum reduction of up to 80.6\%. This results in an average LER reduction of 38.8\%, with a maximum reduction of up to 82.3\%, as shown in Fig.~\ref{fig:parallel_comparison} (right). This highlights the effectiveness of O3LS in optimizing execution schedules with loose scheduling algorithm.

\begin{figure}[h]
    \centering
        \includegraphics[width=1\linewidth]{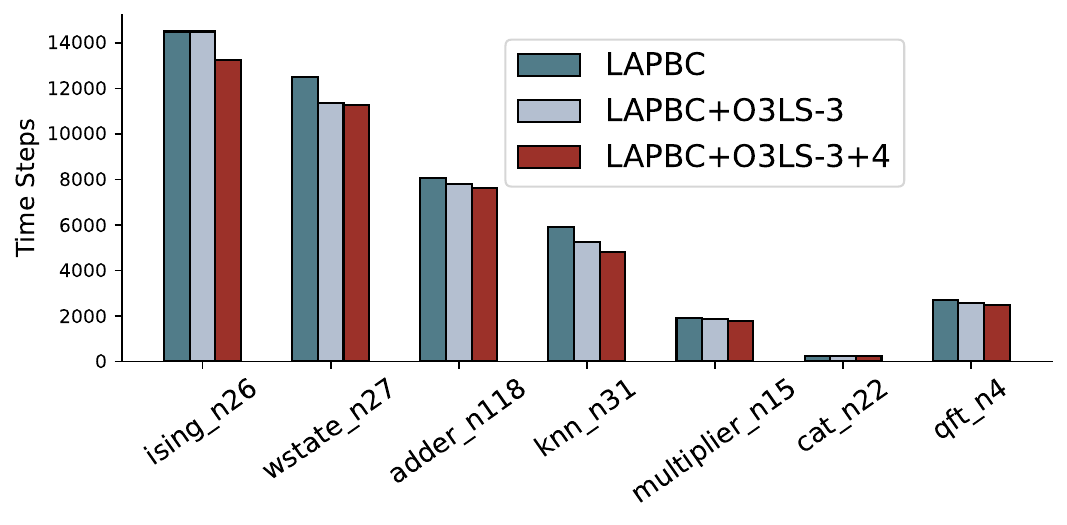}
    \caption{Time step reduction by integrating O3LS with high-parallelism execution strategies.
    }
    \label{fig:fujii_add_ours}
\end{figure}

\textbf{c) Incorporating parallelism into O3LS.} While the highly parallel nature of \texttt{LAPBC} provides advantages for sparse data layouts, O3LS can also be integrated into \texttt{LAPBC} to further reduce time costs. In particular, we focus on several high-parallelism benchmarks where \texttt{LAPBC} is expected to perform well. By incorporating additional modules such as loose scheduling and advanced initial mapping algorithms, we further improve performance and demonstrate the effectiveness of the proposed methods. A detailed breakdown of the results is shown in Fig.~\ref{fig:fujii_add_ours}, where our integrated approach achieves an average improvement of 9.31\%.

\subsection{Initial Mapping Comparison}

\begin{figure}[h]
    \centering
    \includegraphics[width=1\linewidth]{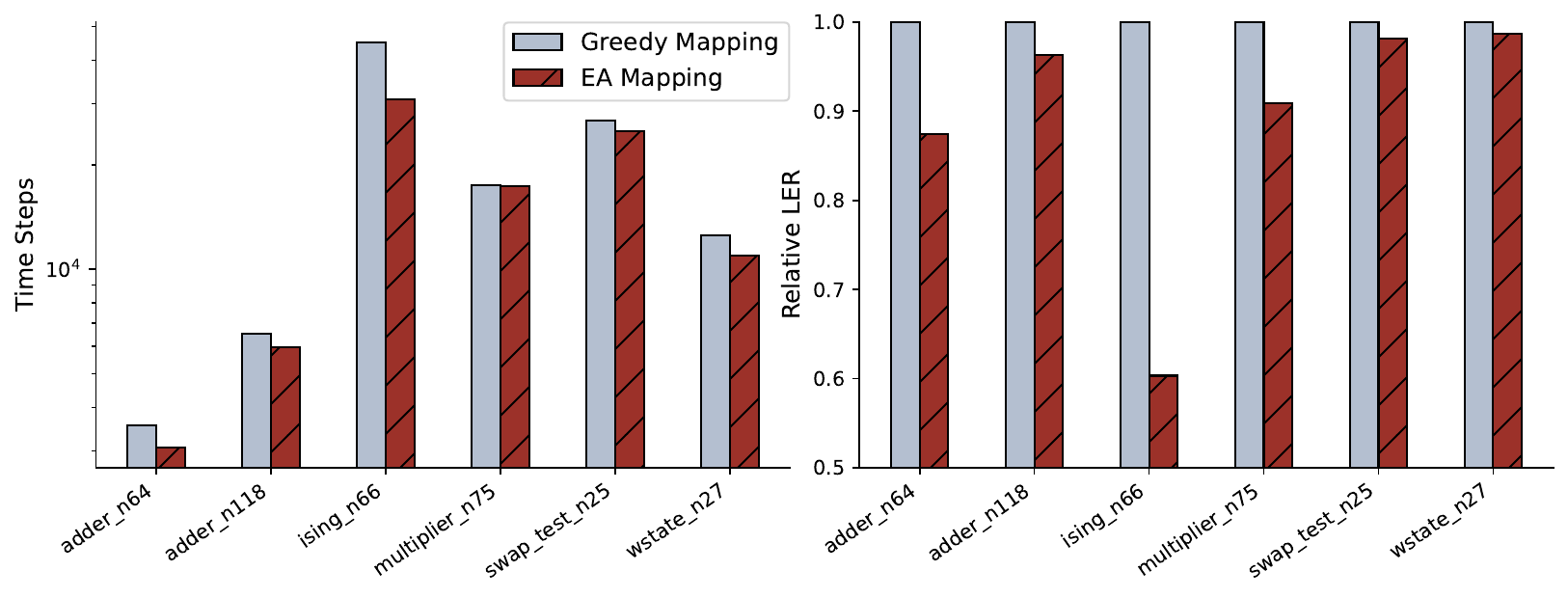}
    \caption{Initial mapping comparison.}
    \label{fig:initial_mapping}
\end{figure}

Furthermore, we analyze the initial mapping methods proposed in O3LS and compare them with the previous greedy mapping approach from~\cite{kan2025sparo}. We use the data layouts generated by O3LS to evaluate these initial mappings. As shown in Fig.~\ref{fig:initial_mapping}, the edge-aware mapping outperforms the previous approach, achieving a time step reduction of 15.0\% and a logical error rate reduction of 8.4\%. These savings primarily result from the edge-aware mapping's tendency to place qubits with higher rotation demands in patches where both X and Z operators are adjacent to an ancilla patch, thereby reducing the need for costly rotations. This strategy is especially effective for the squeezed layouts produced by O3LS.

\subsection{Overall Performance Comparison}

\textbf{a) Overall Performance.} After comparing individual components, we directly compare the overall performance of O3LS with compiler passes including SPC and LAPBC in terms of LER. 
The exact LER results are presented in Fig.~\ref{fig:exact_LER}. Compared to \texttt{SPC}, O3LS suppresses the logical error rate by 43.11\% and 44.98\% on compact and standard layouts, respectively (corresponding to a reduction by roughly half). Notably, in certain cases, O3LS achieves a maximum LER reduction of 93.95\% compared to \texttt{LAPBC}, approaching an order of magnitude error suppression.

\begin{figure*}[t]
    \centering
    \includegraphics[width=1\linewidth]{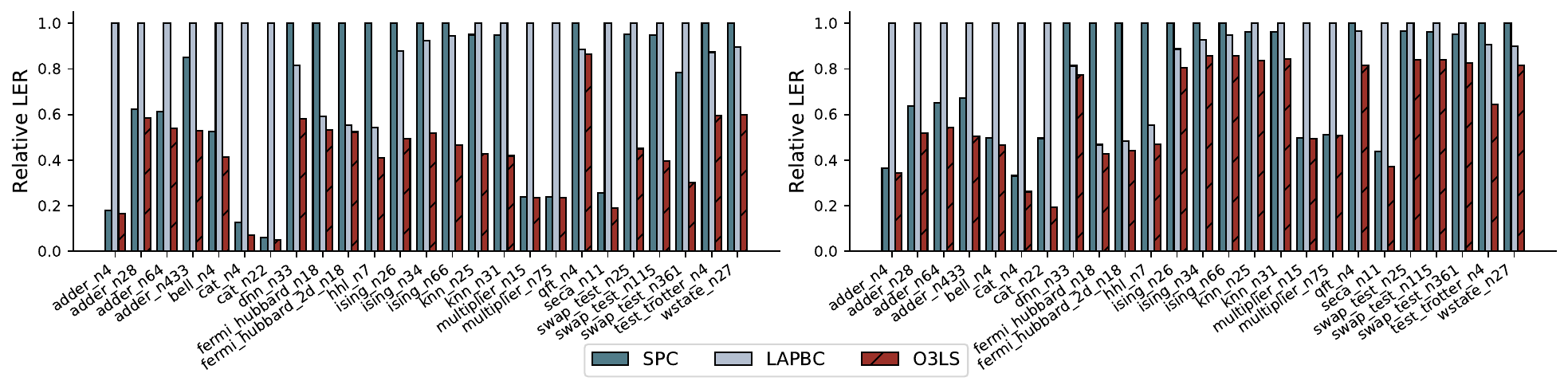}
    \caption{Relative logical error rate comparison (lower is better). O3LS vs. prior compilers with fixed compact (left) or standard (right) data layouts.}
    \label{fig:full_comparision}
\end{figure*}

\begin{figure}[t]
    \centering
    \includegraphics[width=1\linewidth]{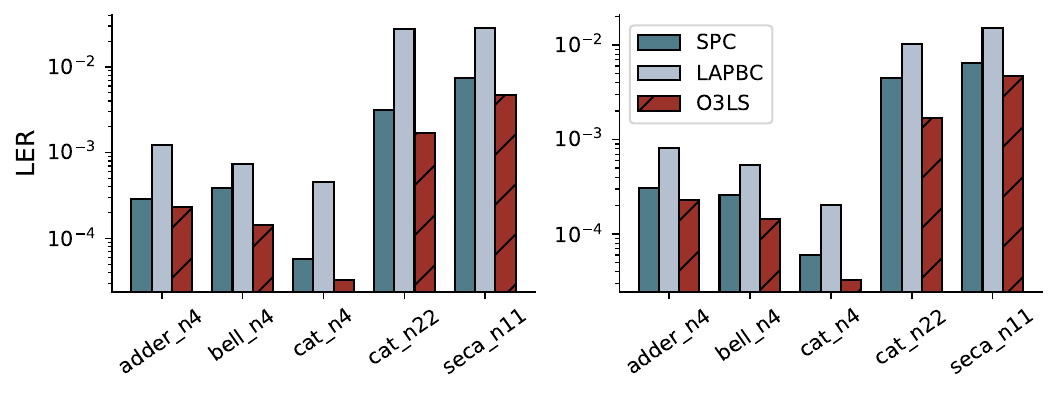}
    \caption{Logical error rate comparison with prior compilers for (left) compact and (right) standard layouts (lower is better).}
    \label{fig:exact_LER}
\end{figure}

Since the simulated surface code distances and the level of improvement vary across applications, we also record the results and normalize them into relative LER values by setting the highest LER for each application as the baseline to clearly show the improvements.
Overall, O3LS outperforms \texttt{SPC}, reducing relative LER by 35.9\% and 21.8\% on compact and standard layouts, respectively. Similarly, compared to \texttt{LAPBC}, O3LS achieves relative LER reductions of 50.9\% and 31.1\%.

Overall, O3LS outperforms previous compiler passes primarily due to all module design choices. First, it identifies the trade-off between scheduling overhead and ancilla path length during layout design, leading to more efficient data layouts. It also applies circuit synthesis techniques to reduce the total number of operations, thereby lowering time overhead associated with rotating data patches in the O3LS-generated layouts. Second, rather than relying on fixed scheduling schemes, O3LS introduces a flexible scheduling strategy that minimize scheduling costs. 
Third, it incorporates effective initial mapping techniques that improve both the routing efficiency.
Together, these strategies not only reduce time overhead but also minimize the need for long ancilla paths, resulting in suppression of LER.

\begin{figure}[h]
    \centering
    \includegraphics[width=1\linewidth]{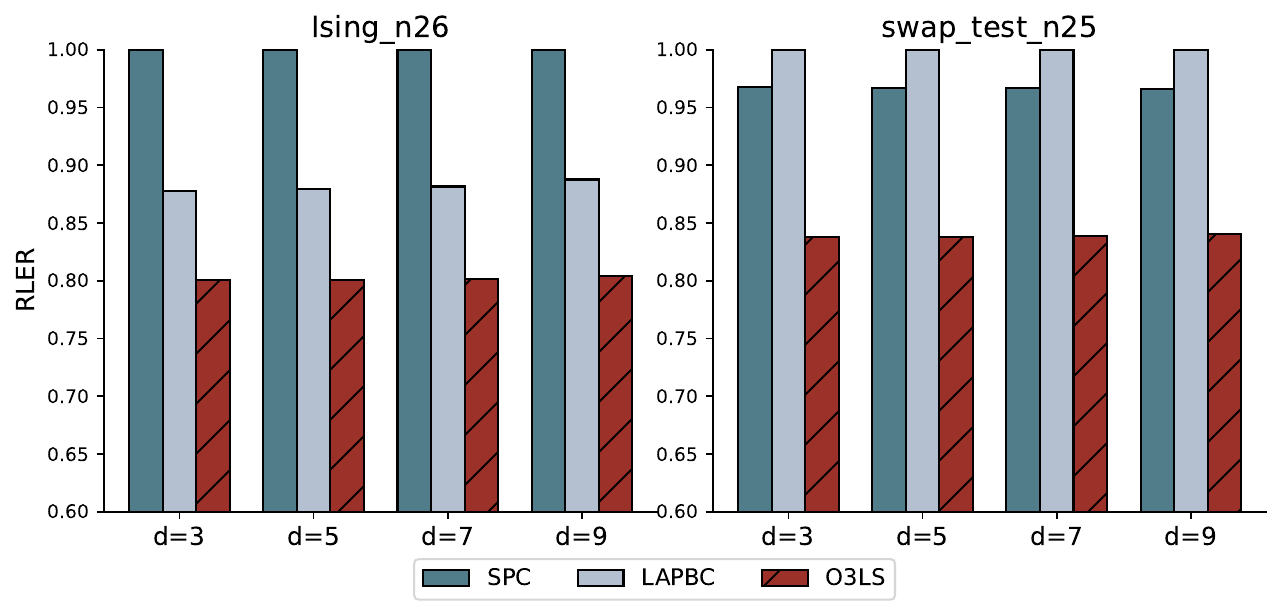}
    \caption{Performance of O3LS scale with code distance $d\in[3,5,7,9]$.}
    \label{fig:sensitivity_analysis_distance}
\end{figure}

\textbf{b) Sensitivity Analysis on Code Distance.} Fig.~\ref{fig:sensitivity_analysis_distance} shows a sensitivity analysis over surface code distance $d \in [3,5,7,9]$. The results show that O3LS consistently outperforms all previous compilers across all tested code distances. For the Ising\_n26 and swap\_test\_n25 benchmark, O3LS consistently achieves over 19.96\% and 13.42\% improvement in RLER compared to \texttt{SPC} and \texttt{LAPBC}, respectively. Crucially, this relative improvement remains stable as the code distance increases. This is because, under the layer-wise accumulation model that treats logical failures as independent rare events~\cite{kan2025sparo}, O3LS optimizes only architectural factors, while the distance-dependent exponential suppression from decoding applies equally to O3LS and the baseline. Consequently, our gains are not tied to any particular code distance.

\begin{figure}[h]
    \centering
    \includegraphics[width=1\linewidth]{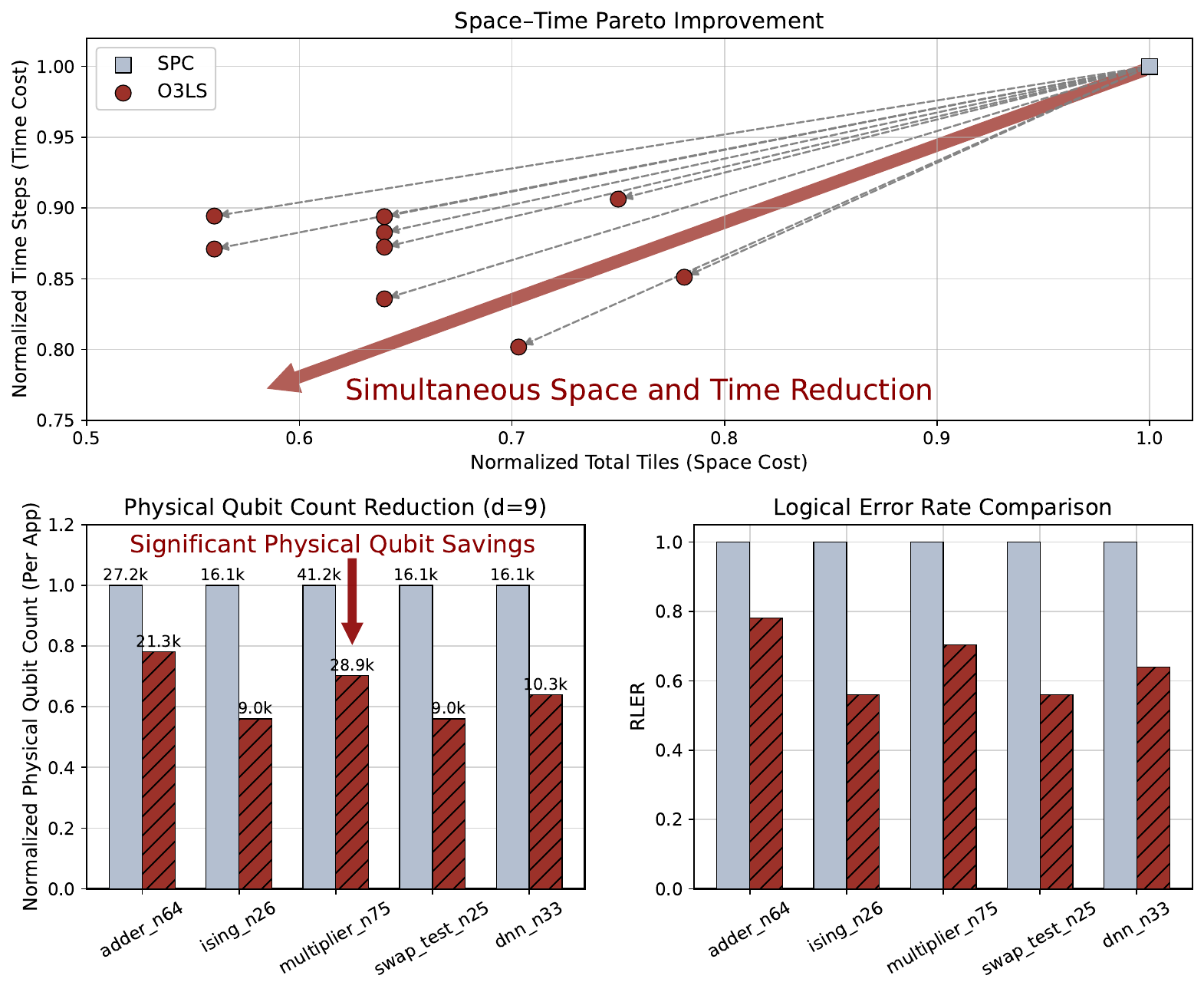}
    \caption{Analysis of resource estimation.}
    \label{fig:resource_estimation}
\end{figure}

\textbf{c) Analysis on Resource Estimation.} After demonstrating robustness across code distances and noise models in improving LER for all benchmarks, Fig.~\ref{fig:resource_estimation} (upper) reports resource estimates for overall time (number of syndrome-measurement cycles) and space (physical qubit count) savings achieved by O3LS relative to SPC. Overall, O3LS reduces space and time simultaneously, delivering an average 23.63\% improvement compared with the prior compiler and thereby suppressing LER (Fig.~\ref{fig:resource_estimation} (lower right)) through automated layout design and scheduling.

Fig.~\ref{fig:resource_estimation} (lower left) further quantifies the space savings using the surface code with $d=9$ as an example. Because each tile corresponds to one surface-code logical qubit, reducing the tile count directly reduces the number of physical qubits. In our benchmarks, O3LS achieves up to a 44\% space reduction, which corresponds to saving roughly 7000 physical qubits. This benefit becomes even more pronounced at larger surface-code distances. This highlights the usefulness of O3LS, which reduces not only time steps but also the physical-qubit requirements for executing fault-tolerant algorithms on hardware.

\subsection{Ablation Study of Compilation Techniques}

\begin{figure*}[t]
    \centering
    \includegraphics[width=1\linewidth]{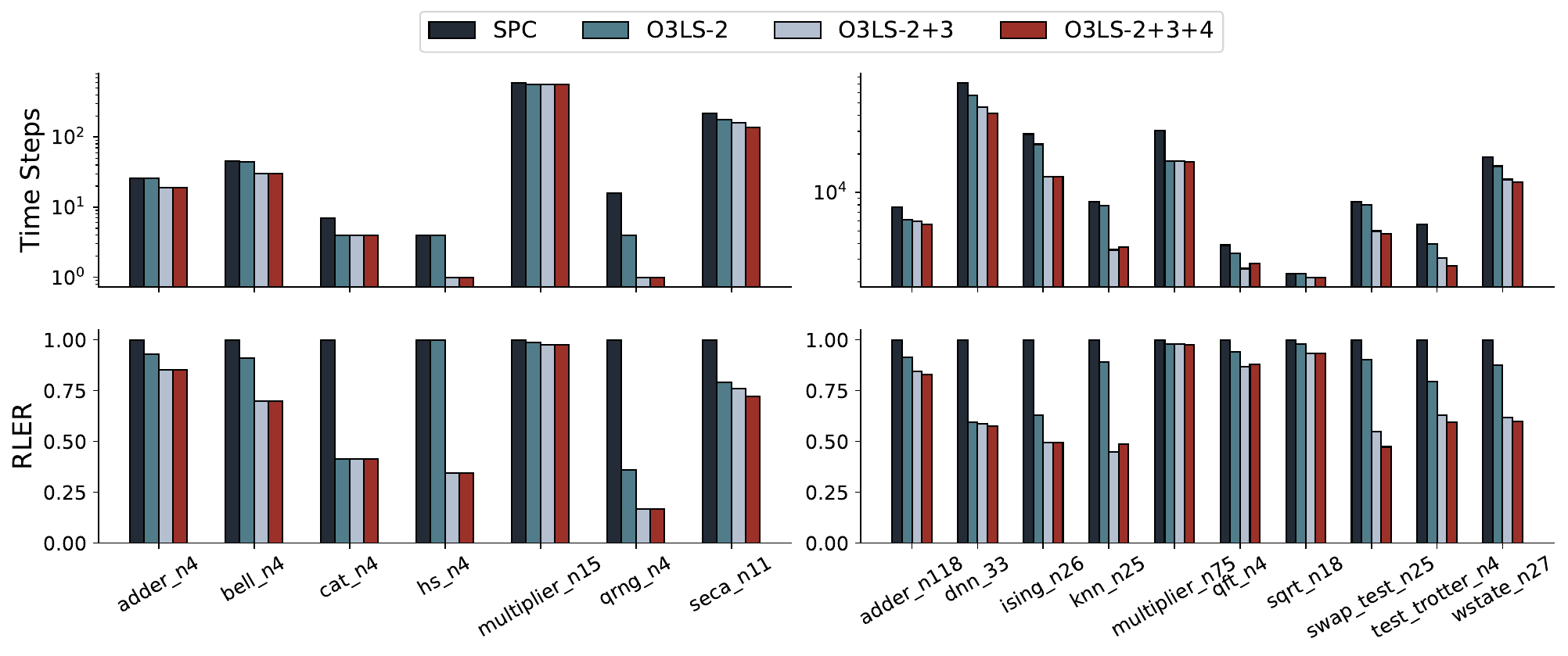}
    \caption{Ablation study of compilation techniques. O3LS-2 refers to the use of Y-synthesis algorithm without additional scheduling methods. O3LS-2+3 incorporates loose scheduling in addition to Y-synthesis, while O3LS-2+3+4 integrates both initial mapping and loose scheduling alongside the Y-synthesis.}
    \label{fig:ablation_study}
\end{figure*}

Fig.~\ref{fig:ablation_study} presents a comparative analysis of the individual modules in O3LS, using the data layout generated by the `O3LS-1' configuration as the test case.
`O3LS-2' refers to the compiler pass that utilizes only the Module 2 Y-synthesis algorithm described in Sec.~\ref{sec:module1}. 
Compared with prior compiler passes, `O3LS-2' improves time steps by 18.33\% and LER by 18.30\%, highlighting the potential of operator cancellation for optimizing circuit execution. With the integration of loose scheduling, `O3LS-2+3' achieves an average improvement of 37.74\% in time steps and 34.34\% in LER, demonstrating the added benefit of loose scheduling. Finally, incorporating initial mapping technique in `O3LS-2+3+4' further improves performance, yielding an average improvement of 38.62\% in time steps and 35.17\% in LER. The results highlight that combining Y-synthesis, loose scheduling, and edge-aware mapping reduces execution time and improves lattice surgery compilation efficiency.

\subsection{Compilation Time Analysis}

\begin{figure}[h]
    \centering
    \includegraphics[width=1\linewidth]{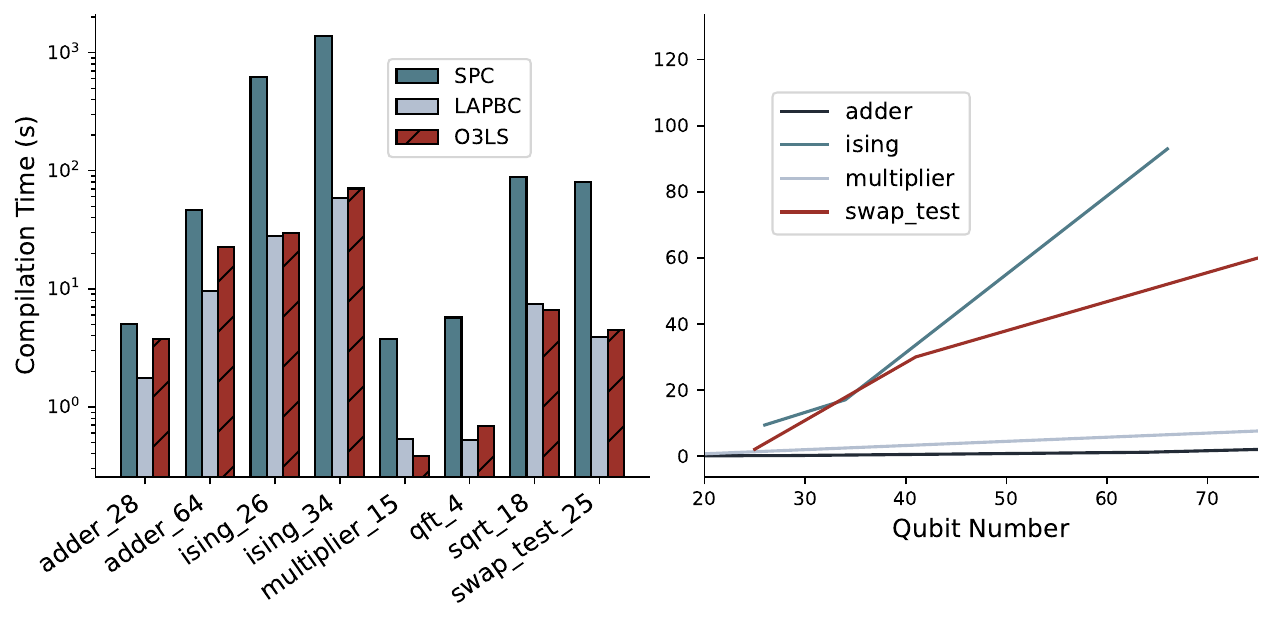}
    \caption{Compilation time analysis. (Left) Comparison with previous compilers. (Right) Scalability with respect to the number of logical qubits. }
    \label{fig:compilation_time}
\end{figure}

We compare the compilation time across different compilers in Fig.~\ref{fig:compilation_time} (left). O3LS achieves faster compilation times than \texttt{SPC} and delivers comparable performance to \texttt{LAPBC}. In some cases, O3LS is slightly slower than \texttt{LAPBC}, which benefits from maximizing parallelism and avoiding the overhead of absorbing Pauli operators into the final measurement. Although O3LS performs explicit Pauli operator transformations, it leverages the O3LS-IR to accelerate this process more effectively than \texttt{SPC}, resulting in compilation times that remain competitive with \texttt{LAPBC}. In Fig.\ref{fig:compilation_time} (right), we also demonstrate the scalability of O3LS with respect to the number of qubits, showing that its compilation time scales polynomially, as analyzed in Section~\ref{sec:compiler_method}.

\subsection{Optimality Analysis}

\begin{figure}[h]
    \centering
    \includegraphics[width=0.95\linewidth]{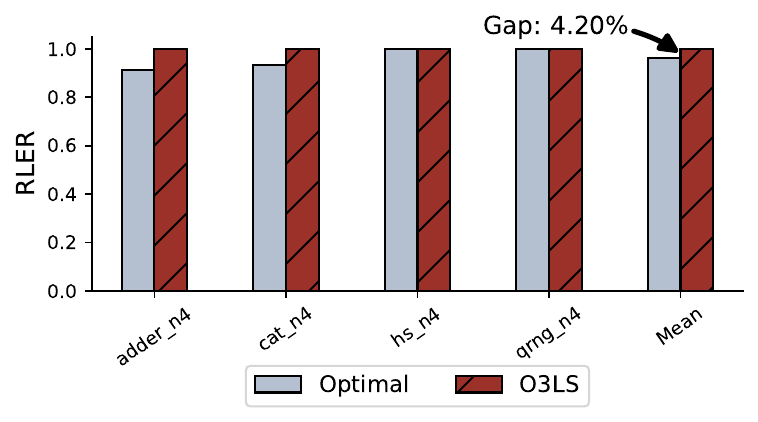}
    \caption{Optimality analysis.}
    \label{fig:optimal_analysis}
\end{figure}

We further conduct an optimality analysis, with the results presented in Fig.~\ref{fig:optimal_analysis}.
Due to the NP-hardness of the problem\cite{herr2017optimization}, our analysis focuses only on small cases, where the optimal LER can be determined through brute-force enumeration.
Overall, O3LS achieves an average gap of just 4.20\% from the optimal, demonstrating the effectiveness of O3LS.

\section{Comparison with Prior Art}

\textbf{a) Lattice Surgery Compilers.} Prior work has explored quadratic assignment~\cite{lao2018mapping} and SAT formulations~\cite{molavi2025dependency} for data layout assignment and scheduling. However, lattice surgery optimization is NP-hard~\cite{herr2017optimization}, requiring scalable solutions. General compilers~\cite{watkins2024high, leblond2023realistic} miss key optimization opportunities in synthesis, mapping, and scheduling. Our work addresses these gaps via improved loose scheduling methods and Y-synthesis algorithms. Meanwhile, methods to enhance parallelism~\cite{beverland2022assessing, beverland2022surface, hamada2024latticesurgery, hirano2025localityaware} show strong results in specific cases. O3LS achieves comparable performance and can integrate these techniques for further gains.

\textbf{b) Lattice Surgery Pipeline.} \cite{wang2024optimizing} proposed TACO to reduce Clifford cost by minimizing Pauli-Z rotations. \cite{chatterjee2025qspellbook} introduced Q-Spellbook for selecting data block layouts and distillation protocols under various strategies. Our work focuses on a different but complementary aspect, and their techniques could be integrated into O3LS to further reduce overall cost.

\textbf{c) Data Layout Design.} \textit{Designs with a similar setting. } In the context of manually designed layouts, \cite{chamberland2022universal} proposed a 4/9 filling layout, while \cite{beverland2022surface} and \cite{beverland2022assessing} introduced 1/4 and 1/2 filling layouts. Although these designs ensure that any logical operation on the target data patches can be executed, they often overlook opportunities for optimizing logical error rates.
In the context of automated layout design, \cite{kan2025sparo} automatically enlarges the underlying data layout based on the analyzed bottleneck.
LaSsynth~\cite{Tan2024LaS} proposes a SAT-based solver that can optimally handle a limited number of qubits and operations. However, its scalability is limited. Our aim is to develop an automated and scalable compiler for finding squeezed layouts.

\textit{Heterogeneous QEC designs. } \cite{stein2023hetarch} introduces a toolbox for heterogeneous quantum architectures on superconducting devices, while \cite{xu2024constant, bravyi2024high, stein2025hetec} propose hybrid approaches that combine surface codes with qLDPC codes, leveraging their complementary strengths by assigning different codes to memory and computation regions. In all cases, surface-code architectures remain central, and our work can potentially offer improved pipelines to enhance their performance.

\textbf{d) Mapping QEC codes into hardware.} \cite{wu2022synthesis} presents a synthesis framework for surface codes on superconducting devices, while \cite{yin2025qecc} extends this to stabilizer code mapping. \cite{leblond2023tiscc, yin2025flexion} explore surface code mapping on trapped-ion devices. These studies primarily focus on the lower layer of mapping QEC codes to physical hardware. Our work is orthogonal to these efforts, and integrating both layers has the potential to further reduce logical error rates.

\section{Conclusion}

We present O3LS, a compiler that suppresses logical error rates by optimizing both space and time overhead for lattice surgery operations. It produces data layouts that minimize space overhead while maintaining time costs comparable to those of sparser layouts, thereby supporting the goal of reducing logical error rates. O3LS achieves this through loose scheduling, Y operator synthesis, and initial mapping tailored to the proposed layout architecture. The numerical results demonstrate that O3LS could outperform prior works in terms of error rates, time costs, and qubit resource overhead.

\section*{Acknowledgment}
We would like to thank the anonymous reviewers for their helpful feedback and suggestions.

\newpage 
\bibliographystyle{IEEEtranS}
\bibliography{publish_ver/ref.bib}

\end{document}